\documentclass[apj,numberedappendix,tighten]{emulateapj}

\pdfoutput=1

\usepackage{amsmath}

\PassOptionsToPackage{breaklinks=true}{hyperref}

\usepackage[lf,mathtabular,footnotefigures]{MinionPro}
\makeatletter
 \def\Mn@Text@Family{MinionPro-TLF}
 
\makeatother

\def\***#1{\textbf{\boldmath\textsf{***#1***}}}

\renewcommand{\deg}{\ensuremath{^\circ}}

\newcommand{\fgas}{f_{\text{gas}}}

\newcommand{\TX}{\ensuremath{T_{\!X}}}
\def\Tx2{\ensuremath{T_{\!X,2}}}
\newcommand{\YX}{\ensuremath{Y_{\mkern -1mu X}}}
\newcommand{\LX}{\ensuremath{L_{X}}}
\def\M500{\ensuremath{M_{500}}}
\def\R500{\ensuremath{r_{500}}}

\newcommand{\Mtot}{\ensuremath{M_{\text{tot}}}}
\newcommand{\Mgas}{\ensuremath{M_{\text{gas}}}}
\newcommand{\Msun}{\ensuremath{M_\odot}}
\newcommand{\LCDM}{$\Lambda$CDM}
\def\400d{400d}

\newcommand\OmegaK{\ensuremath{\Omega_{k}}}
\newcommand\OmegaX{\ensuremath{\Omega_{X}}}
\newcommand\OmegaM{\ensuremath{\Omega_{\rm M}}}
\newcommand\OmegaL{\ensuremath{\Omega_\Lambda}}
\newcommand\OmegaMh{\ensuremath{\Omega_{\rm M}h}}

\newcommand\DeltaR{\ensuremath{\Delta_{\frak R}}}
\newcommand\zCMB{\ensuremath{z_{\rm CMB}}}

\newcommand\wo{\ensuremath{w_0}}
\newcommand\wz{\ensuremath{w(z)}}
\newcommand\wa{\ensuremath{w_a}}

\slugcomment{Submitted 5/12/08; Revised 10/3/08; Accepted 10/29/2008}
\journalinfo{The Astrophysical Journal, in press (692, 2009 February
  10); 	arXiv:0812.2720}

\shorttitle{\emph{CHANDRA}\/ CLUSTER COSMOLOGY PROJECT.\quad III.}
\shortauthors{VIKHLININ ET AL.}

\begin{document}

\title{\emph{Chandra}\/ Cluster Cosmology Project III: COSMOLOGICAL
  PARAMETER CONSTRAINTS}

\author{
A.~Vikhlinin\altaffilmark{1,2},
A.~V.~Kravtsov\altaffilmark{3},
R.~A.~Burenin\altaffilmark{2},
H.~Ebeling\altaffilmark{4},
W.~R.~Forman\altaffilmark{1},
A.~Hornstrup\altaffilmark{5},
C.~Jones\altaffilmark{1},
S.~S.~Murray\altaffilmark{1},
D.~Nagai\altaffilmark{6},
H.~Quintana\altaffilmark{7},
A.~Voevodkin\altaffilmark{2,8}
}

\altaffiltext{1}{Harvard-Smithsonian Center for Astrophysics, 
                 60 Garden Street, Cambridge, MA 02138}
\altaffiltext{2}{Space Research Institute (IKI), Profsoyuznaya 84/32,
                 Moscow, Russia}
\altaffiltext{3}{Dept.\ of Astronomy and Astrophysics,
  Kavli Institute for Cosmological Physics, Enrico Fermi Institute,
  University of Chicago, Chicago, IL 60637}
\altaffiltext{4}{Institute for Astronomy, University of Hawaii, 2680
Woodlawn Drive, Honolulu, HI 96822}
\altaffiltext{5}{National Space Institute, Technological University of
  Denmark, Juliane Maries Vej 30, DK-2100 Copenhagen, Denmark}
\altaffiltext{6}{Department of Physics and Yale Center for Astronomy \&
  Astrophysics, Yale University, New Haven, CT 06520}
\altaffiltext{7}{Departamento de Astronomia y Astrofisica, 
                 Pontificia Universidad Catolica de Chile, Casilla 306, 
                 Santiago, 22, Chile}
\altaffiltext{8}{Los Alamos National Laboratory, Los Alamos, NM 87545}

\begin{abstract}
  \emph{Chandra} observations of large samples of galaxy clusters
  detected in X-rays by \emph{ROSAT} provide a new, robust determination
  of the cluster mass functions at low and high redshifts. Statistical
  and systematic errors are now sufficiently small, and the redshift
  leverage sufficiently large for the mass function evolution to be used
  as a useful growth of structure based dark energy probe. In this
  paper, we present cosmological parameter constraints obtained from
  \emph{Chandra} observations of 37~clusters with $\langle
  z\rangle=0.55$ derived from 400~deg$^2$ \emph{ROSAT} serendipitous
  survey and 49 brightest $z\approx 0.05$ clusters detected in the
  All-Sky Survey.  Evolution of the mass function between these
  redshifts requires $\OmegaL>0$ with a $\sim 5\sigma$ significance, and
  constrains the dark energy equation of state parameter to
  $\wo=-1.14\pm0.21$, assuming constant $w$ and flat universe.  Cluster
  information also significantly improves constraints when combined with
  other methods. Fitting our cluster data jointly with the latest
  supernovae, WMAP, and baryonic acoustic oscillations measurements, we
  obtain $w_0=-0.991\pm0.045$ (stat) $\pm0.039$ (sys), a factor of 1.5
  reduction in statistical uncertainties, and nearly a factor of 2
  improvement in systematics compared to constraints that can be
  obtained without clusters.  The joint analysis of these four datasets
  puts a conservative upper limit on the masses of light neutrinos,
  $\sum m_\nu<0.33$~eV at 95\% CL.  We also present updated measurements
  of $\OmegaMh$ and $\sigma_8$ from the low-redshift cluster mass
  function.
\end{abstract}
\keywords{cosmology: observations, cosmological parameters, dark matter ---
  clusters: general --- surveys}

%===================================================================
\section{Dark Energy and Cluster Mass Function} 
\label{sec:intro} 

Recent accelerated expansion of the Universe detected in the Hubble diagram
for distant type~Ia supernovae is one of the most significant discoveries of
the past 10 years \citep{1999ApJ...517..565P,1998AJ....116.1009R}. The
acceleration can be attributed to the presence of a significant energy
density component with negative pressure, hence the phenomenon is commonly
referred to as Dark Energy. For a recent review of the dark energy discovery
and related theoretical and observational issues, see
\cite*{2008arXiv0803.0982F} and references therein. Perhaps the simplest
phenomenological model for dark energy is non-zero Einstein's cosmological
constant. The supernovae data indicated (and other cosmological datasets now
generally agree) that a cosmological constant term currently dominates
energy density in the Universe.

The next big question is whether Dark Energy really is the cosmological
constant. The properties of dark energy are commonly characterized by
its equation of state parameter, $w$, defined as $p=w\rho$, where $\rho$
is the dark energy density and $p$ is its pressure. A cosmological
constant in the context of General Relativity corresponds to a
non-evolving $w=-1$. It is proposed that departures from the
cosmological constant model should be sought in the form of observed $w$
being either $\ne -1$, or evolving with redshift.  Combination of
supernovae, cosmic microwave background, and baryonic acoustic
oscillations data currently constrain $|1+w|<0.15$ at 95\% CL
\citep{2008arXiv0803.0547K}.  Observational signatures of such
deviations of $w$ from $-1$ are very small, and hence the measurements
are prone to systematic errors. For example, variations of $w$ between
$-1$ and $-0.9$ change fluxes of $z=0.75$ supernovae in a flat universe
with $\OmegaM=0.25$ by only $0.03$~magnitudes.  Therefore, it is
crucially important that the dark energy constraints at this level of
accuracy are obtained from combination of several independent
techniques.  This not only reduces systematics but also improves
statistical accuracy by breaking degeneracies in the cosmological
parameter constraints.

One of the methods that has been little used so far is evolution in the
number density of massive galaxy clusters. Evolution of the cluster mass
function traces (with exponential magnification) growth of linear density
perturbations. Growth of structure and distance-redshift relation are
similarly sensitive to properties of dark energy, and also are mutually
highly complementary methods \citep[e.g.,][]{2003MNRAS.346..573L}.  Mapping
between the linear power spectrum and cluster mass function relies on the
model for nonlinear gravitational collapse.  This model is now calibrated
extensively by $N$-body simulations (see \S\,\ref{sec:theory}). The cluster
mass function models also use additional assumptions (e.g., that the mass
density is dominated by cold dark matter in the recent past, and that the
fluctuations have Gaussian distribution). However, corrections due to
reasonable departures from these assumptions are negligible compared to
statistical uncertainties in the current samples (we discuss these issues
further in \S\,\ref{sec:theory}).  It is important also that the theory of
nonlinear collapse is insensitive to the background cosmology. For example,
the same model accurately describes the relation between the linear power
spectrum and cluster mass function in the $\OmegaM=1$, $\OmegaL=0$,
low-density $\OmegaM=0.3$, $\OmegaL=0$, and ``concordant'' $\OmegaM=0.3$,
$\OmegaL=0.7$ cosmologies \citep{2001MNRAS.321..372J}.

Fitting cosmological models to the real cluster mass function measurements
uses not only growth of structure but also the distance-redshift information
because observed properties for objects of the same mass generally depend on
the distance. Therefore, constraints on $w$ derived from the cluster mass
function internally make a combination of growth of structure and distance
based cosmological tests, and thus potentially can be very accurate and
competitive with any other technique \citep[e.g.,][]{2006astro.ph..9591A}.

% Ultimately, comparison of the measurements based on growth of structure and
% with the purely geometrical methods can be used as a test of whether the
% accelerated expansion is caused by dark energy or departures of gravity
% theory from general relativity \***{REF}.

Previous attempts to use evolution of the cluster mass function as a
cosmological probe were limited by small sample sizes and either poor
proxies for the cluster mass (e.g., the total X-ray flux) or inaccurate
measurements (e.g., temperatures with large uncertainties). Despite
these limitations, reasonable constraints could still be derived on
$\OmegaM$ \citep[e.g.,][]{2001ApJ...561...13B,2004ApJ...609..603H}.
However, constraints on the dark energy equation of state from such
studies are weak.  For example, \citet{2004ApJ...609..603H} derived the
best-fit $w=-0.42$, only marginally inconsistent with $w=-1$, using the
temperature function of the \emph{Einstein} Medium Sensitivity Survey
clusters; \citet{2007arXiv0709.4294M} determine $w=-1.4\pm0.55$ with a
larger sample of distant clusters \citep[MACS survey,
see][]{2001ApJ...553..668E} but using the X-ray luminosity as a mass
proxy.

The situation with the cluster mass function data has been dramatically
improved in the past two years. A large sample of sufficiently massive
clusters extending to $z\sim 0.9$ has been derived from \emph{ROSAT}
PSPC pointed data covering 400~deg$^2$
\citep[][Paper~I~hereafter]{2007ApJS..172..561B}.  Distant clusters from
the \400d{} sample were then observed with \emph{Chandra}, providing
high-quality X-ray data and much more accurate total mass indicators.
\emph{Chandra} coverage has also become available for a complete sample
of low-$z$ clusters originally derived from the \emph{ROSAT} All-Sky
Survey. Results from deep \emph{Chandra} pointings to a number of
low-$z$ clusters have significantly improved our knowledge of the outer
cluster regions and provided a much more reliable calibration of the
$\Mtot$ vs.\ proxy relations than what was possible before. On the
theoretical side, improved numerical simulations resulted in better
understanding of measurement biases in the X-ray data analysis
\citep{2007ApJ...655...98N,2006MNRAS.369.2013R,2007arXiv0708.1518J}.
Even more importantly, results from these simulations have been used to
suggest new, more reliable X-ray proxies for the total mass
\citep{2006ApJ...650..128K}. We discuss all this issues in the previous
paper \citep[Paper~II hereafter]{vikhlinin08}. The cluster mass
functions derived in this paper are reproduced in Fig.\,\ref{fig:mfun}.
Overall, these results are an important step forward in providing
observational foundation for cosmological work with the cluster mass
functions.

\defcitealias{2007ApJS..172..561B}{Paper~I}

\defcitealias{vikhlinin08}{Paper~II}

In this work, we present cosmological constraints from the data discussed in
\citetalias{vikhlinin08}. The cosmological information contained in the
cluster mass function data and relevant to dark energy constraints can be
approximately separated into 3 quasi-independent components:

(1)\hspace*{0.5em}Changes in the comoving number density at a fixed mass
threshold constrain a combination of the perturbations growth factor and
relative distances between low and high-$z$ samples; this by itself is a
dark energy constraint (\S\ref{sec:w0}).

(2)\hspace*{0.5em}The overall normalization of the observed mass function
constrains the amplitude of linear density perturbations at $z\approx0$,
usually expressed in terms of the $\sigma_8$ parameter.  Statistical and
systematic errors in the $\sigma_8$ measurement are now sufficiently small,
and the ratio of $\sigma_8$ and the amplitude of the CMB fluctuations power
spectrum gives the total growth of perturbations between $z\approx1000$ and
$z=0$ --- a second powerful dark energy constraint
(\S\,\ref{sec:other:data}).

(3)\hspace*{0.5em}The slope of the mass function measures $\OmegaM\times h$;
this by itself is not a dark energy probe but can be used to break
degeneracies present in other methods.

Our dark energy constraints were derived for the following cases.
Assuming constant $w$ and flat universe, we measure $\wo=-1.14\pm0.21$
using only cluster data (i.e., evolution of the mass function between
our two redshift samples) and the HST prior on $h$
(\S\,\ref{sec:wo:clusters}). Combining cluster and WMAP data, we obtain
$\wo=-1.08\pm0.15$ but ($\wo$ is constrained much more tightly for a
fixed $\OmegaM$ (\S\,\ref{sec:wo:clusters+cmb}). Finally, adding cluster
data to the joint supernovae + WMAP + BAO constraint, we obtain
$\wo=-0.991\pm0.045$ (\S\,\ref{sec:wo:clusters+others}), significantly
reducing statistical and especially systematic
(\S\,\ref{sec:w:systematics}) uncertainties compared to the case without
clusters. A large fraction of the extra constraining power comes from
contrasting $\sigma_8$ with normalization of the CMB power spectrum;
this procedure is sensitive to non-zero mass of light neutrinos.
Allowing for $m_\nu>0$, we obtain a new conservative upper limit $\sum
m_\nu<0.33$~eV (95\% CL) while still improving the $w_0$ measurement
relative to the SN+WMAP+BAO-only case ($\wo=-1.02\pm0.055$,
\S\,\ref{sec:mnu}). Adding clusters also improves equation of state
constrains for evolving $w$ in flat universe (\S\,\ref{sec:w0-wa}) and
constant $w$ in non-flat universe (\S\,\ref{sec:w0-Ok})

The paper is organized as follows. We start with a short summary of cluster
data and systematic uncertainties (\S\,\ref{sec:data}), discuss issues
relevant for computing theoretical mass function models
(\S\,\ref{sec:theory}) and describe our fitting procedure
(\S\,\ref{sec:fitting}). We then discuss constraints that can be obtained
from low-redshift mass function only ($\OmegaMh$ in \S\,\ref{sec:OmegaMh}
and $\sigma_8$ in \S\,\ref{sec:s8}). We then consider as an example
constraints from the cluster evolution in non-flat \LCDM{} model (i.e., $w$
fixed at $-1$); $\OmegaL>0$ is required with $\sim 5\sigma$ confidence
(\S\,\ref{sec:O-L}). Constraints on the dark energy equation of state are
considered in \S\S\,\ref{sec:w0}--\ref{sec:w:general}. Systematic errors are
discussed in \S\,\ref{sec:w:systematics}.

\section{Summary of the cluster data and systematic uncertainties}
\label{sec:data}
\label{sec:systematics:in:data}

\begin{figure}
\centerline{\includegraphics[width=\columnwidth]{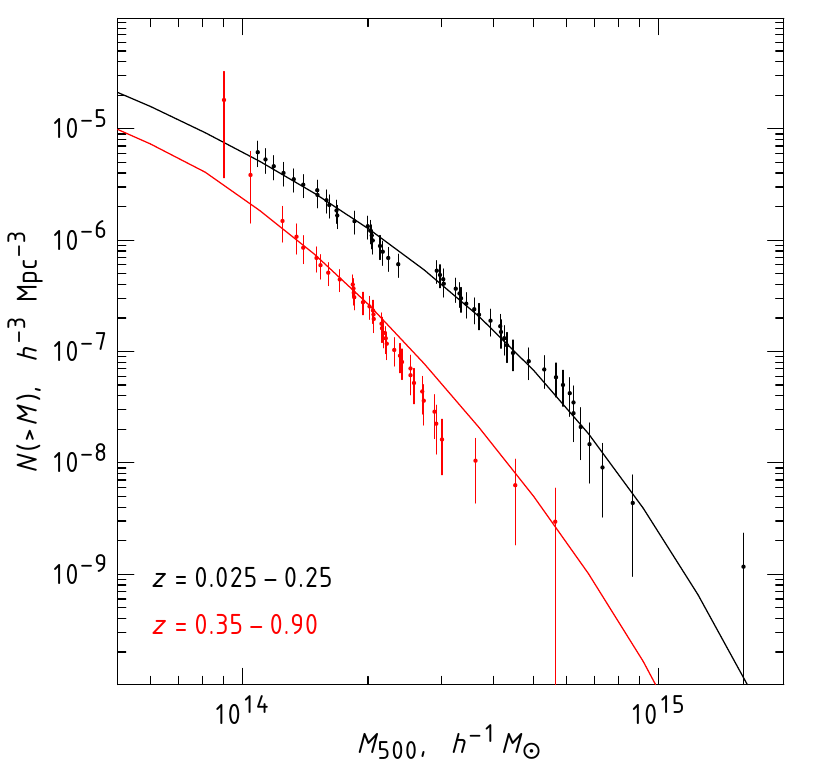}}
\caption{Estimated mass functions for our cluster samples computed for the
  $\OmegaM=0.25$, $\OmegaL=0.75$, $h=0.72$ cosmology.  Solid lines show the
  mass function models (weighted with the survey volume as a function of $M$
  and $z$), computed for the same cosmology with only the overall
  normalization, $\sigma_8$, fitted. The deficit of clusters in the distant
  sample near $M_{500}=3\times10^{14}\,h^{-1}\,\Msun$ is a marginal
  statistical fluctuation --- we observe 4 clusters where 9.5 are expected,
  a $2\sigma$ deviation (cf.\ Fig.\,17 in \citetalias{vikhlinin08}).}
\label{fig:mfun}
\end{figure}

This work is based on two cluster samples, originally compiled from
\emph{ROSAT} X-ray surveys (see \citetalias{vikhlinin08} for a complete
description of the sample selection and data analysis). The low-redshift
sample includes the 49 highest-flux clusters detected in the All-Sky
Survey at Galactic latitudes $|b|>20\deg$ and $z>0.025$. The effective
redshift depth of this sample is $z<0.15$. The high-redshift sample
includes 37 $z>0.35$ objects detected in the 400d survey, with an
additional flux cut applied; the redshift depth of this sample is
$z\approx0.9$.  All the low and high-$z$ clusters were later observed
with \emph{Chandra}, providing good statistical precision spatially
resolved spectral data thus yielding several high-quality $\Mtot$
estimators for each object.  The combined cluster sample is a unique,
uniformly observed dataset. The volume coverage and effective mass
limits at low and high redshifts are similar (see the estimated mass
functions in Fig.\,\ref{fig:mfun}).

Because of the sufficiently high quality of the \emph{Chandra} data, we
employ advanced data analysis techniques going well beyond simple flux
estimates and $\beta$-model fits commonly used in earlier studies.
Cosmological cluster simulations has been used to test for the absence of
significant observational biases in reconstructing the basic cluster
parameters \citep{2007ApJ...655...98N}. Using these simulations, we also
tested which of the X-ray observables are best proxies for the total cluster
mass \citep{2006ApJ...650..128K,2007ApJ...655...98N} and concluded that the
best three are the average temperature, $\TX$, measured in the annulus
$[0.15-1]\,r_{500}$ (thus excluding the central region often affected by
radiative cooling and sometimes, by AGN activity in the central galaxy); the
intracluster gas mass integrated within $r_{500}$; and the combination of
the two, $\YX=\TX\times\Mgas$.  These parameters are low-scatter proxies of
the total mass (in particular, $\YX$, and $\Mgas$ is only slightly worse).
Simulations and available data show that the scaling of these proxies with
$\Mtot$, including the redshift dependence, is very close to predictions of
the simple self-similar model.  In a sense, even though we use advanced
numerical simulations which include multiple aspects of the cluster physics
to test $\Mtot$ vs.\ proxy relations, the role of simulations is limited to
providing small corrections to predictions of very basic and hence reliable
theory. Application of these corrections as well as practical considerations
for deriving $\TX$, $\Mgas$, and $\YX$ from the real data are discussed in
\citetalias{vikhlinin08}. 

\begin{figure*}
\centerline{\includegraphics[width=\columnwidth]{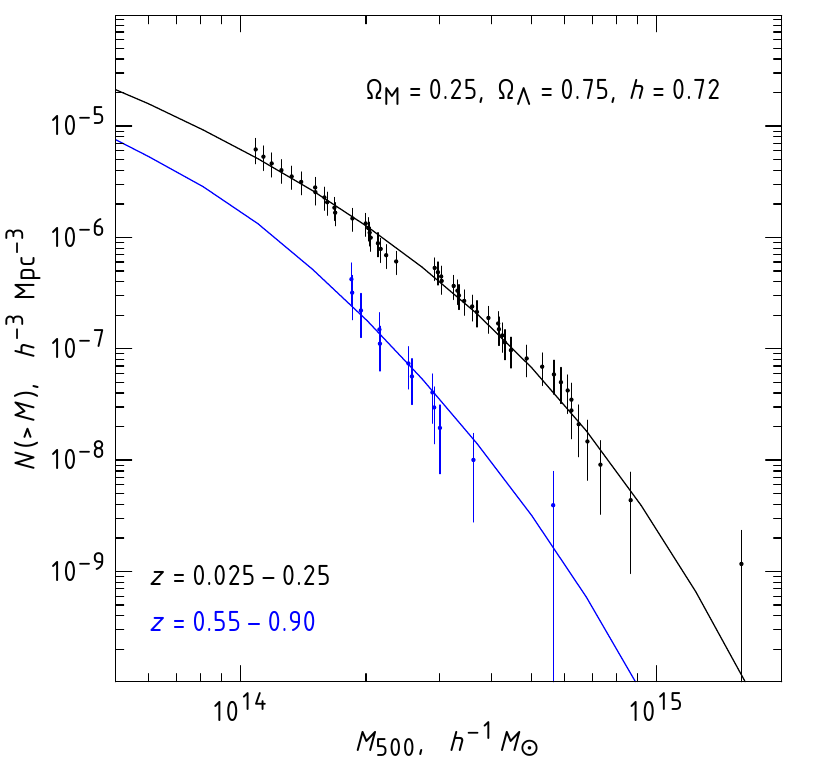}\includegraphics[width=\columnwidth]{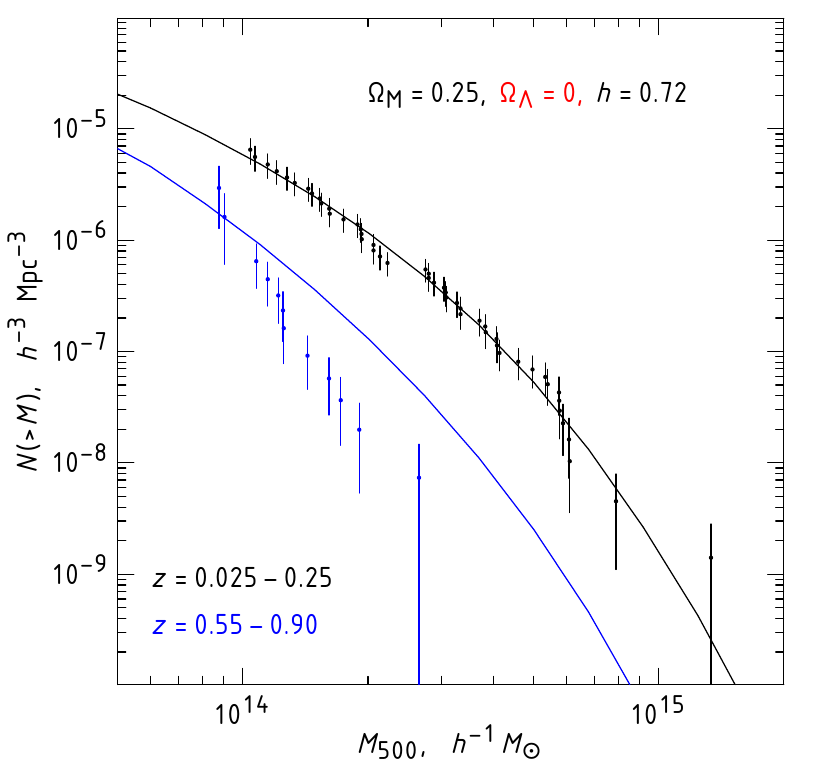}}
\caption{Illustration of sensitivity of the cluster mass function to the
  cosmological model. In the left panel, we show the measured mass
  function and predicted models (with only the overall normalization at
  $z=0$ adjusted) computed for a cosmology which is close to our
  best-fit model. The low-$z$ mass function is reproduced from
  Fig.\,\ref{fig:mfun}, which for the high-$z$ cluster we show only the
  most distant subsample ($z>0.55$) to better illustrate the effects. In
  the right panel, both the data and the models are computed for a
  cosmology with $\OmegaL=0$. Both the model and the data at high
  redshifts are changed relative to the $\OmegaL=0.75$ case. The
  measured mass function is changed because it is derived for a
  different distance-redshift relation. The model is changed because the
  predicted growth of structure and overdensity thresholds corresponding
  to $\Delta_{\rm crit}=500$ are different. When the overall model
  normalization is adjusted to the low-$z$ mass function, the predicted
  number density of $z>0.55$ clusters is in strong disagreement with the
  data, and therefore this combination of $\OmegaM$ and $\OmegaL$ can be
  rejected.}
\label{fig:mfun:lambda-nolambda}
\end{figure*}

\citetalias{vikhlinin08} also presents an observational calibration of the
$\Mtot$ vs.\ proxy relations using an extremely well-observed sample of
low-$z$ clusters. This discussion is crucial for understanding the
systematic uncertainties in our cluster mass function measurements, and we
urge interested readers to consult \citetalias{vikhlinin08}. Table~4 there
gives a summary of the main sources of systematic uncertainties in the
derived cluster mass functions. They can be separated into three
quasi-independent components. First is the uncertainty in calibration of the
absolute cluster mass scale by \emph{Chandra} hydrostatic mass estimates in
a sample of dynamically relaxed, well-observed low-$z$ clusters
\citep{2006ApJ...640..691V}; the level of this uncertainty (9\%) is
estimated from comparison of \emph{Chandra} masses with two recent weak
lensing studies \citep{2007MNRAS.379..317H,2008A&A...482..451Z}. Second is
uncertainties related to possible departures from standard evolution in
$\Mtot-\TX$, $\Mtot-\Mgas$, and $\Mtot-\YX$ relations. This uncertainty
($\sim 5-6\%$ between $z=0$ and $z=0.5$) was estimated from general
reliability of numerical models of the cluster formation and from the
magnitude of corrections that had to be applied to the data (see \S\,4 in
\citetalias{vikhlinin08} for details). The last major source of uncertainty
is evolution in the $\LX-\Mtot$ relation, affecting computations of the 400d
survey volume coverage; this uncertainty is mostly measurement in nature
because we derive the $\LX-\Mtot$ relation internally from the same cluster
set. Its effect is negligible for the high-$M$ end of the mass function and
becomes comparable to Poisson errors for low-$M$ clusters. A representative
compilation of the effects of $\LX-\Mtot$ uncertainties on the $V(M)$
function is presented in Fig.\,15 of \citetalias{vikhlinin08}.

The general reliability of our analysis is greatly enhanced by using
independent, high-quality X-ray indicators of the total cluster mass ---
$\TX$, $\Mgas$, $\YX$. Since the masses estimated from these proxies depend
differently on the distance to the object, the high-$z$ mass functions
estimated with different proxies should agree only if the assumed background
cosmology is correct. In principle, this can be used as an additional source
of information for the distance-redshift relation and folded into the
cosmological fit. However, this method is nearly equivalent to the
$\fgas(z)$ test, which is more reliably carried which is more relibaly
carried out using direct hydrostatic mass estimates in relaxed clusters
\citep{2008MNRAS.383..879A}, and therefore we ignore this information.
Instead, we use the agreement between different proxies observed for the
best-fit cosmology as a comforting indication that there are no serious
errors in our results.

\section{Summary of theory}
\label{sec:theory}

In the current paradigm of structure formation, galaxy clusters form via
gravitational collapse of matter around large peaks in the primordial
density field \citep{1984ApJ...284L...9K,1986ApJ...304...15B}. Their
abundance and spatial distribution in a comoving volume will thus depend
on the statistical properties of the initial density field, such as
gaussianity\footnote{We note however, that the current constraints on
  non-gaussianity from the CMB anisotropy measurements imply that the
  expected effects on clusters are small \citep{2007MNRAS.382.1261G}.}
and power spectrum (and hence the cosmological parameters that determine
it), and could depend on the details of non-linear amplitification of
the density perturbations by gravity. Indeed, semi-analytic models based
on the linear primordial density field and a simple ansatz describing
non-linear gravitational collapse of density peaks
\citep{1974ApJ...187..425P,1991ApJ...379..440B,1998ApJ...500...14L,2001MNRAS.323....1S}
have proven to be quite successful in describing results of direct
cosmological simulations of structure formation
\citep[e.g.,][]{1999ApJ...517L...5L,2001MNRAS.323....1S,2001MNRAS.321..372J}.
The accuracy of the existing models, however, is limited and over the
last several years the abundance of collapsed objects was calibrated by
fitting appropriate fitting function to the results of direct
cosmological simulations
\citep{2001MNRAS.321..372J,2002ApJ...573....7E,2006ApJ...646..881W}. The
fitting functions are expressed in the so-called universal
form\footnote{In the sense that the same function and parameters could
  be used to predict halo abundance for different redshifts and
  cosmologies.} as a function of the variance of the density field on
the mass scale $M$. The fact that such universal expressions exist
implies that there is a direct link between the the linearly evolving
density field and cluster abundance.

In our analysis we use the most recent accurate calibration of the halo
mass function by \citet{2008arXiv0803.2706T}, which provides fitting
formulas for halo abundance as a function of mass, defined in spherical
apertures enclosing overdensities similar to the mass we derive from
observational proxies for the observed clusters. The
\citeauthor{2008arXiv0803.2706T}\ fitting formulas are formally accurate
to better than $5\%$ for the cosmologies close to the concordance
\LCDM{} cosmology and for the mass and redshift range of interest in our
study; at this level, the theoretical uncertainties in the mass function
do not contribute significantly to the systematic error budget.
Although the formula has been calibrated using dissipationless $N$-body
simulations (i.e.  without effects of baryons), the expected effect of
the internal redistribution of mass during baryon dissipation on halo
mass function are expected to be $<5\%$ \citep{2008ApJ...672...19R} for
a realistic fraction of baryons that condenses to form galaxies.

Similarly to \citet{2001MNRAS.321..372J} and \citet{2006ApJ...646..881W},
the Tinker et al. formulas for the halo mass function are presented as a
function of variance of the density field on a mass scale $M$. The variance,
in turn, depends on the linear power spectrum of the cosmological model,
$P(k)$, which we calculate as a product of the initial power law spectrum,
$k^n$, and the transfer function for the given mixture of CDM and baryons,
computed using the analytic approximations of \citet{1999ApJ...511....5E}.
This analytic approximation is accurate to better than 2\% for a wide range
of cosmologies, including cosmologies with non-negligible neutrino
contributions to the total matter density.

Our default analysis assumes that neutrinos have a negligibly small
mass. The only component of our analysis that could be affected by this
assumption is when we contrast the low-redshift value of $\sigma_8$
derived from clusters with the CMB power spectrum normalization. This
comparison uses evolution of purely CDM+baryons power spectra. The
presence of light neutrinos affects the power spectrum at cluster
scales; in terms of $\sigma_8$, the effect is roughly proportional to
the total neutrino density, and is $\approx 20\%$ for $\sum m_\nu =
0.5$~eV \citep[we calculate the effect of neutrinos using the transfer
function model of][]{1999ApJ...511....5E}. Stringent upper limits on the
neutrino mass were reported from comparison of the WMAP and Ly-$\alpha$
forest data, $\sum m_\nu<0.17$~eV at 95\% CL
\citep{2006JCAP...10..014S}. If neutrino masses are indeed this low,
they would have no effect on our analysis.  However, possible issues
with modeling of the Ly-$\alpha$ data have been noted in the literature
\citep[see, e.g., discussion in \S\,4.2.8 of][]{2008arXiv0803.0586D} and
so we experiment also with neutrino masses outside the Ly-$\alpha$
forest bounds (\S\,\ref{sec:mnu}).

\begin{deluxetable*}{p{3cm}cp{4.5cm}cp{4.5cm}}
  \tablecaption{Cosmological constraints from X-ray cluster data\label{tab:par:clusters}}
  \tablehead{
    \colhead{Parameter} &
    \colhead{Value} &
    \colhead{Determined by} &
    \colhead{Systematic errors} &
    \colhead{Dominant source}
    \\
    \colhead{}  &
    \colhead{} & 
    \colhead{} &
    \colhead{} &
    \colhead{of systematic uncertainties}
%     \\
%     \colhead{(1)}  &
%     \colhead{(2)} &
%     \colhead{(3)} &
%     \colhead{(4)} &
%     \colhead{(5)}
  }
  \startdata
  $\Omega_Mh$\dotfill & $0.184\pm0.024$ 
               &\raggedright Shape of the local mass function,
               \S\,\ref{sec:OmegaMh} 
               & $\pm 0.027$
               &{\raggedright Slope of the $L-M$ relation.}\\
  $\Omega_M$\dotfill & $0.255\pm0.043$ 
               &\raggedright Shape of the local mass function plus HST
               prior on $h$,
               \S\,\ref{sec:OmegaMh} 
               & $\pm 0.037$
               &{\raggedright Slope of the $L-M$ relation.}\\
  $\sigma_8(\Omega_M/0.25)^{0.47}$\dotfill & $0.813\pm0.013$ 
               & \raggedright Normalization of the local mass function,
               \S\,\ref{sec:s8}
               & $\pm 0.024$ 
               &{\raggedright Absolute mass calibration at $z=0$.}\\
  $\Omega_M$\dotfill & $0.34\pm0.08$ 
               & \raggedright Evolution of the $\TX$-based mass function,
               \S\,\ref{sec:O-L}
               & $\pm 0.055$
               &{\raggedright Evolution of the $M-T$ relation}
  \enddata
\end{deluxetable*}

\section{Fitting procedure}
\label{sec:fitting}

We obtain parameter constraints using the likelihood function computed
on a full grid of cosmological parameters affecting cluster observables
(and also those for external datasets). The relevant parameters for the
cluster data are those that affect the distance-redshift relation, as
well as the growth and power spectrum of linear density perturbations:
$\OmegaM$, $\OmegaL$, $w$ (dark energy equation of state parameter),
$\sigma_8$ (linear amplitude of density perturbations at the
$8\,h^{-1}$~Mpc scale at $z=0$), $h$, tilt of the primordial
fluctuations power spectrum, and potentially, the non-zero rest mass of
light neutrinos. This is computationally demanding and we describe our
approach below.

The computation of the likelihood function for a single combination of
parameters is relatively straightforward.  Our procedure (described in
\citetalias{vikhlinin08}) uses the full information contained in the
dataset, without any binning in mass or redshift, takes into account the
scatter in the $\Mtot$ vs.\ proxy relations and measurement errors, and so
on. We should note, however, that since the measurement of the $\Mgas$ and
$\YX$ proxies depends on the assumed distance to the cluster, the mass
functions must be re-derived for each new combination of the cosmological
parameters that affect the distance-redshift relation --- $\OmegaM$, $w$,
$\OmegaL$, etc.  Variations of $h$ lead to trivial rescalings of the mass
function and do not require re-computing the mass estimates.  Computation of
the survey volume uses a model for the evolving $\LX-\Mtot$ relation
\citepalias[see \S\,5 in][]{vikhlinin08}, which is measured internally from
the data and thus also depends on the assumed $d(z)$ function. Therefore, we
refit the $\LX-\Mtot$ relation for each new cosmology and recompute $V(M)$.
Sensitivity of the derived mass function to the background cosmology is
illustrated in Fig.\,\ref{fig:mfun:lambda-nolambda}. The entire procedure,
although equivalent to full reanalysis of the \emph{Chandra} and
\emph{ROSAT} data, can be organized very efficiently if one stores the
derived $\rho_g(r)$ and $T(r)$ computed in some reference cosmology. It
takes $\approx 20$~sec on a single CPU to re-estimate all masses, refit the
$\LX-\Mtot$ relation, and recompute volumes for each new combination of the
cosmological parameters.

The next step is to compute, for each combination of $\OmegaM$, $\OmegaL$
etc., the likelihood function on a grid of those parameters which do not
affect the distance-redshift relation. In our case, these are $\sigma_8$,
$h$, and when required, the power spectrum tilt or neutrino mass. The
cluster datad are extremely sensitive to $\sigma_8$ and so we need a fine
grid for this parameter. Fortunately, the mass function codes compute the
mass functions for different values of $\sigma_8$ with other parameters
fixed at almost no extra expense.  The sensitivity of the cluster data to
$h$ and tilt is much weaker, therefore the likelihood can be computed on a
coarse grid for these parameters and then interpolated.

With the acceleration strategies outlined above, it took us $\sim
9600$~CPU-hours (or 20 days using multiple workstations) to compute the
cluster likelihood functions on full parameter grids for several generic
models (non-flat \LCDM, constant dark energy equation of state in a flat
universe, constant $w$ with non-zero neutrino mass, linearly evolving $w$ in
flat universe, constant $w$ in non-flat universe). Alternatively, simulating
the Markov chains \citep{2002PhRvD..66j3511L} with sufficient statistics for
all these cases would require approximately the same computing time.

After the cluster likelihood function was computed, we also computed
$\chi^2$ for external cosmological datasets --- WMAP (5-year results),
Baryonic Acoustic Oscillations, and Supernovae~Ia bolometric distances.
Since we basically use analytic Gaussian priors for these datasets (see
\S\,\ref{sec:other:data} below), these computations are fast and can be
made on a fine parameter grid. We also use a Gaussian prior for the
Hubble constant, $h=0.72\pm0.08$, based on the results from the HST Key
Project \citep{2001ApJ...553...47F}. This prior is important only when
the constraints from the shape of the mass function
(\S\,\ref{sec:OmegaMh}) come into play and when external cosmological
datasets are not used in the constraints.  When fitting the cluster
data, we also keep the absolute baryon density fixed at the best-fit
WMAP value, $\Omega_bh^2=0.0227$ \citep{2008arXiv0803.0586D}.
This parameter slightly affects the calculation of the linear power
spectrum \citep{1998ApJ...496..605E}.  When we add the WMAP
information to the total constraints, we marginalize the WMAP
likelihood component over this parameter. If not stated otherwise, our
cosmological fits also assume a primordial density fluctuation power
spectrum with $n=0.95$ \citep{2007ApJS..170..377S}. Our results are
completely insensitive to variations of $n$ within the WMAP
measurement uncertainties and even to setting $n=1$.

Once the combined likelihood as a function of cosmological parameters is
available, we use the quantity $-2\ln L$, whose statistical properties
are equivalent to the $\chi^2$ distribution \citep{1979APJ...228..939C},
to find the best fit parameters and confidence intervals.

In addition to statistical uncertainties, we also consider different
sources of systematics. We do not include systematic errors in the
likelihood function but instead refit parameters with the relations
affected by systematics varied within the estimated $1\sigma$
uncertainties. This approach allows as not only to estimate how the
confidence intervals are expanded from combination of all systematic
errors, but also to track the most important source of uncertainty for
each case. A full analysis of systematic errors is presented in
\S\,\ref{sec:w:systematics} for the case of constraints on constant $w$
in a flat universe; in other cases the systematic uncertainties
contribute approximately the same fraction of the total error budget. We
also verified that in the constant $w$ case, our method of estimating
the systematic errors produces the results which are very close to the
more accurate procedure using the Markov chain analysis.

\section{Constraints from the shape of the local mass function:
  $\OmegaMh$}
\label{sec:OmegaMh}

The shape of the cluster mass function reflects the shape of the linear
power spectrum in the relevant range of scales, approximately
$10\,h^{-1}$~Mpc in our case. This shape, for a reasonable range of
parameters in the CDM cosmology is controlled \citep{1986ApJ...304...15B}
mostly by the quantity $\OmegaMh$. It is useful to consider constraints on
this combination separately because they are nearly independent of the rest
of the cosmological parameters we are trying to measure with the cluster
data.

Fixing the primordial power spectrum index to the WMAP value, $n=0.95$,
the fit to the local mass function\footnote{Including the high-redshift
  data, we obtain a consistent value, $\OmegaMh=0.198\pm0.022$. Combined
  with the HST prior on $h$, this leads to a measurement of
  $\OmegaM=0.275\pm0.043$.  However, using the high-$z$ data makes the
  $\OmegaMh$ constraints dependent on the background cosmology and
  therefore we prefer to base this measurement only on the local mass
  function. Also, we use the $\YX$-based mass estimates for this and
  $\sigma_8$ analyses. The other observables, $\TX$ or $\Mgas$, give
  essentially identical results, because all of them were normalized
  using the same set of low-$z$ clusters \citepalias[see][for
  details]{vikhlinin08}. The difference between mass proxies is only
  important for the measurements based on the evolution of the high-$z$
  mass function (\S\,\ref{sec:O-L}).} gives $\OmegaMh = 0.184 \pm 0.024$
(purely statistical 68\% CL uncertainties). The best fit value is
degenerate with the assumed primordial power spectrum index, and the
variation approximately follows the relation $\Delta \OmegaMh =
-0.31\Delta n$.  The variations of $n$ within the range constrained by
the WMAP data, $\pm0.015$, lead to negligibly small changes in our
derived $\OmegaMh$.

An additional source of statistical uncertainty is that related to the
derivation of the $L-M$ relation, since we derive this relation from the
same set of clusters. Uncertainties in the $L-M$ relation are translated
into those of the survey volume and hence the cluster mass function.
Most of our cosmological constraints are primarily sensitive to the
cluster number density near the median mass of the sample. This median
mass, the $V(M)$ uncertainties are small compared to statistics (see \S6
in \citetalias{vikhlinin08}). The $\OmegaMh$ determination, however, is
based on the relative number density of clusters near the high and low
mass ends of the sample. Since the volume is a fast-decreasing function
at low $M$'s, the $V(M)$ variations are important. The most important
parameter of the $L-M$ relation in our case is the power law slope,
$\alpha$ (see eq.\,20 in \citetalias{vikhlinin08}). Variations of
$\alpha$ within the errorbars ($\pm 0.14$) of the best fit value lead to
changes in the derived $\OmegaMh$ of $\pm 0.027$. Adding this in
quadrature to the formal statistical errors quoted above, we obtain a
total uncertainty of $\pm0.035$. We have verified that other sources of
systematics in the $\OmegaMh$ determination are much less important than
those related to the $L-M$ relation.

In principle, a non-zero mass of light neutrinos has some effect on the
perturbation power spectrum at low redshifts. We checked, however, that
their effect on the \emph{shape} of the cluster mass function is
negligible for any $\sum m_\nu$ within the range allowed by the CMB data
\citep{2008arXiv0803.0547K}. Therefore, neutrinos do not affect our
results on $\OmegaMh$.

\begin{figure}
\centerline{\includegraphics[width=\columnwidth]{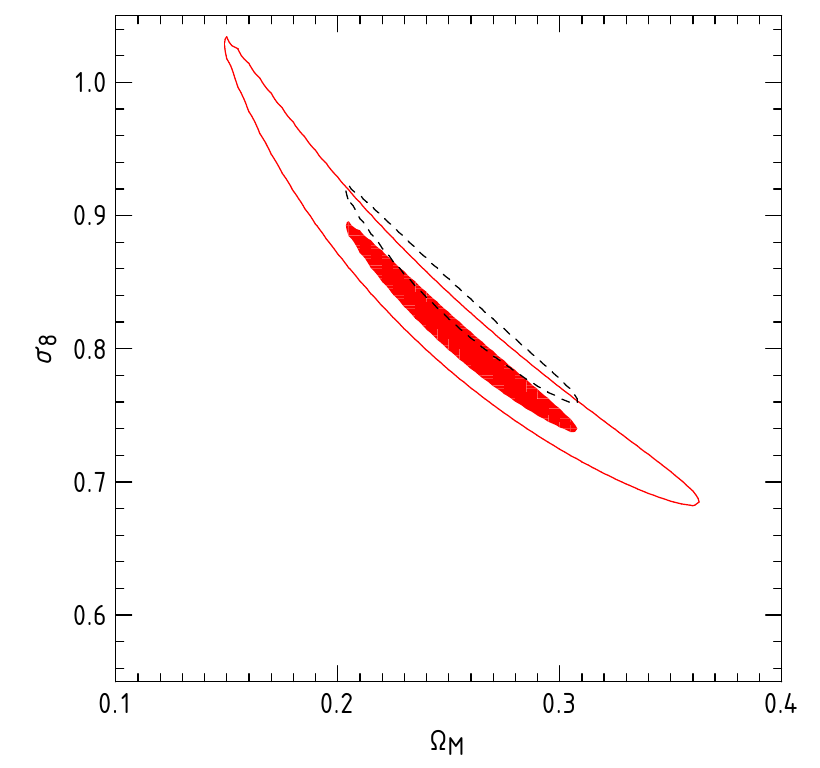}}
\caption{Constraints on the $\sigma_8$ and $\Omega_M$ parameters in a
  flat \LCDM{} cosmology from the total (both low and high-redshift)
  cluster sample. The inner solid region corresponds to $-2\Delta \ln
  L=1$ from the best-fit model (indicates the 68\% CL intervals for one
  interesting parameter, see footnote~\ref{fn:chi2}) and the solid
  contour shows the one-parameter 95\% CL region ($-2\Delta \ln L=4$).
  The dashed contour shows how the inner solid confidence region is
  modified if the normalization of the absolute cluster mass vs.\
  observable relations is changed by $+9\%$ (our estimate of the
  systematic errors).  }
\label{fig:OmegaM-h}
\end{figure}

Our determination of $\OmegaMh=0.184\pm0.035$ compares well with the
previous measurements using cluster data and galaxy power spectra.  Of
the previous cluster results especially noteworthy is the work of
\citet{2003A&A...398..867S} whose constraints are based not only on the
shape of the mass function but also on the clustering of low-$z$
clusters. Their value is $\OmegaMh=0.239\pm0.056$ (errors dominated by
uncertainties in the conversion of cluster X-ray luminosities into mass;
this source of uncertainty is avoided in our work by using high-quality
X-ray mass proxies). $\OmegaMh$ is measured accurately also by galaxy
redshift surveys. The results from the 2dF and SDSS surveys are $\OmegaM
h = 0.178\pm0.016$ and $0.223 \pm 0.023$, respectively \citep[][---we
rescaled to $n=0.95$ their best fit values reported for
$n=1$]{2005MNRAS.362..505C,2004PhRvD..69j3501T}. The individual
errorbars in galaxy survey results are smaller than those from the
cluster data; however, a recent work by \citet{2007ApJ...657..645P}
suggests that the previous galaxy redshift results may be affected by
scale-dependent biases on large scales. Indeed, there is a tension
between the SDSS and 2dF values at $\simeq 90\%$ CL and the difference
is comparable to the errorbars of our measurement.

The cluster results can be improved in the future by extending the range
of the mass function measurements. Not only can this improve statistical
errors in the mass function measurements but it can also improve the
accuracy of the $L-M$ relation, a significant source of uncertainty in
our case.  We note that it is more advantageous to increase statistics
in the high-$M$ range than to extend the mass function into the galaxy
group regime. In addition to greater reliability of the X-ray mass
estimates in the high-$M$ systems, the surveys become dominated by
cosmic variance approximately below the lower mass cut in our sample
\citep[the cosmic variance is estimated in \S7.1 of
\citetalias{vikhlinin08} using the prescription
of][]{2003ApJ...584..702H}.

Combined with the HST prior on the Hubble constant, our constraint on
$\OmegaMh$ becomes a measurement for the matter density parameter,
$\OmegaM=0.255\pm0.043$ (stat) $\pm0.037$ (sys), where systematic errors
are also dominated by the slope of the $L-M$ relation. This agrees
within the errors with other independent determinations, such as a
combination of BAO and CMB acoustic scales, $\OmegaM=0.256\pm0.027$
\citep{2007ApJ...657...51P}, and a combination of gas fraction
measurements in massive clusters with the average baryon density from
Big Bang Nucleosynthesis, $\OmegaM=0.28\pm0.06$
\citep{2008MNRAS.383..879A}. It also agrees with another independent
measurement based on our data, $\OmegaM=0.30\pm0.05$ from evolution of
the cluster temperature function, see (\S\,\ref{sec:O-L} below).

% \begin{verbatim}
% All data:
% h=0.707 +- 0.101    [assuming OmegaM=0.28]
% Delta h = -1.1 Delta n
%    This is equivalent to 
% OmegaM*h = 0.198 +- 0.028
% or Delta OmegaM*h = -0.31*Delta n
%     -- note that the CMB uncertainties
%        on n make a negligible
%        contribution to OmegaM*h
% 
% Using only z=0 data:
% h = 0.656 +- 0.11
% Delta h = -1.1 Delta n
% or
% OmegaM*h = 0.184 +- 0.031
% Delta OmegaM*h = -0.31 * Delta n
% 
% with h=0.72+-0.08, this becomes a constraint on OmegaM: Omega=0.256 +- 0.052
% 
% \end{verbatim}
% 
% $L-M$ slope related uncertainties:
% \begin{verbatim}
% For alpha=1.75 [+1sigma] and h=0.65, best fit n is 1.176
% for alpha=1.6  [default]                      n = 1.018
% Delta n = 0.158 -> Delta Omega*h = 0.033 (if Delta Omega*h = 0.21*Deltan
% as in Voevodkin et al.)
% \end{verbatim}

\begin{figure}
\centerline{\includegraphics[width=\columnwidth]{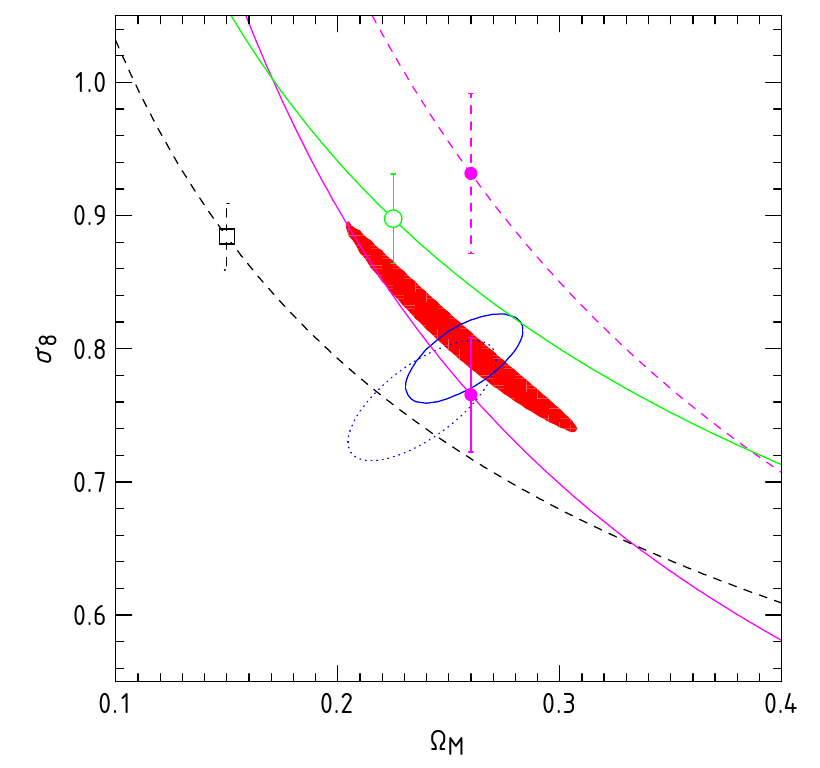}}
\caption{Comparison with other $\sigma_8$ measurements. Solid region is
  our 68\% CL region reproduced from Fig.\,\ref{fig:OmegaM-h} (this and
  all other confidence regions correspond to $\Delta\chi^2=1$, see
  footnote~\ref{fn:chi2} on page~\pageref{fn:chi2}). Blue contours show
  the WMAP 3 and 5-year results from \citet{2007ApJS..170..377S} and
  \citet{2008arXiv0803.0586D} (dotted and solid contours, respectively).
  For other measurements, we show the general direction of degeneracy as
  a solid line and a 68\% uncertainty in $\sigma_8$ at a representative
  value of $\Omega_M$. Filled circles show the weak lensing shear
  results from \citet{2006ApJ...647..116H} and
  \citet{2008A&A...479....9F} (dashed and solid lines, respectively).
  Open circle shows results from a cluster sample with galaxy dynamics
  mass measurements \citep{2007ApJ...657..183R}. Finally, open square
  shows the results from \citet[approximately the lower bound of
  recently published X-ray cluster measurements]{2002ApJ...567..716R}.}
\label{fig:OmegaM-h-comp}
\end{figure}

\section{Constraints from the normalization of the cluster mass
  function: $\sigma_8-\OmegaM$}
\label{sec:s8}

The normalization of the cluster mass function is exponentially
sensitive to $\sigma_8$, the amplitude of linear perturbations at the
length scale $8\,h^{-1}$~Mpc, approximately corresponding to the cluster
mass scale \citep{1990ApJ...351...10F}. Measuring this parameter with
the cluster data has been a popular topic of research, especially using
statistics of X-ray clusters \citep[and many others
thereafter]{1990ApJ...351...10F,1991ApJ...372..410H,1992ApJ...386L..33L,1993MNRAS.262.1023W}.
The strong sensitivity of the predicted cluster number density to
$\sigma_8$ makes the determination of this parameter relatively
insensitive to the details of the sample selection.  Historically,
different studies using very different cluster catalogs yielded similar
results, if the data were analyzed uniformly.  Determination of
$\sigma_8$ is more sensitive to calibration of the absolute mass scale.
For example \citet{2003MNRAS.342..163P} show that if $\Mtot$ for a fixed
value of $\TX$ is varied by a factor of 1.5, $\sigma_8$ derived from the
local cluster temperature function is changed by
$\Delta\sigma_8\approx0.13$. Smaller biases are introduced if the
effects of scatter in deriving the mass-luminosity relation are
neglected resulting in incorrect computations of the survey volume
\citep{2006ApJ...648..956S}. Our present work includes advances in both
of these areas and thus it is worth presenting an updated measurement of
$\sigma_8$.

Determination of $\sigma_8$ from the cluster abundance data usually
shows a strong degeneracy with the $\OmegaM$ parameter, typically,
$\sigma_8\propto\OmegaM^{-0.6}$ \citep[e.g.,][]{2002ApJ...578L..95H}.
The nature of this degeneracy is that the mass function determines the
rms amplitude of fluctuations at the given $\Mtot$ scale. The
corresponding length scale is a function of $\OmegaM$ ($M\sim \OmegaM
l^3$) and thus the derived $\sigma_8$ depends also on $\OmegaM$ and more
weakly on the local slope of the linear power spectrum \citep[see
discussion in][]{1993MNRAS.262.1023W}
% \footnote{We note, however, that
%   the exact form of the degeneracy also depends on the mass proxy used
%   in the analysis. For example, \citet{2004ApJ...601..610V} measured
%   $\sigma_8$ from the number density of clusters as a function of their
%   gas mass and assuming that the ratio of $\Mgas/\Mtot$ is close to the
%   cosmic value, $\Omega_b/\OmegaM$ and its value is fixed by other data.
%   Conversion of the cluster gas mass to the corresponding length scale
%   does not depend on $\OmegaM$ (because the average baryon density in
%   the Universe is fixed by other cosmological data), and so the
%   \citeauthor{2004ApJ...601..610V} determination of $\sigma_8$ is not
%   degenerate with $\OmegaM$.}. 
We need, therefore, to constrain $\sigma_8$ and $\OmegaM$ jointly. We
used a grid of parameters of the flat\footnote{ The assumption of
  flatness (and background cosmology in general) has a minor effect on
  determination of $\sigma_8$ because the measurement is dominated by
  the low-redshift sample. However, we note that when we use the
  $\sigma_8$ information in the dark energy constraints
  (\S\,\ref{sec:w0} and thereafter), we do not use the results from this
  section directly.  When we fit $w$, $\sigma_8$ is effectively
  re-measured from the cluster data for each background cosmology.}
\LCDM{} model ($\OmegaM$, $h$, $\sigma_8$), and computed the cluster
likelihood using the mass function for the local sample. We then add the
Hubble constant prior (\S\,\ref{sec:fitting}), and marginalized the
combined likelihood over $h$.

The results are shown in Fig\,\ref{fig:OmegaM-h}\footnote{The contours
  in these and subsequent figures correspond to the 95\% CL region for
  one interesting parameter ($\Delta\chi^2=4$). The inner solid region
  corresponds to $\Delta\chi^2=1$. This choice is made to facilitate
  quick estimates of the single-parameter uncertainty intervals directly
  from the plots. The total extent of the $\Delta\chi^2=1$ region in
  either direction is a good estimate for the 1-parameter 68\% CL
  interval \citep{1976A&A....52..307C}. Similarly, the width of this
  region is a 68\% CL interval assuming that the second parameter is
  fixed.
  \label{fn:chi2}}. For a fixed $\OmegaM$, the value of $\sigma_8$ is
constrained to within $\pm0.012$ (statistical).  The degeneracy between
$\sigma_8$ and $\OmegaM$ can be accurately described as
$\sigma_8=0.813(\OmegaM/0.25)^{-0.47}$. The $\OmegaM$ range along this
line is constrained by the shape of the local mass function combined
with the HST prior on the Hubble constant (\S\,\ref{sec:OmegaMh}).
Including the high-redshift data, we obtain very similar results. For
example, for $\Omega_M=0.25$, the total sample gives
$\sigma_8=0.803\pm0.0105$, to be compared to $\sigma_8=0.813\pm0.012$
from low-$z$ clusters only. This implies that the $\sigma_8$ measurement
is dominated by the more accurate local cluster data, as expected.

Systematic errors of the $\sigma_8$ measurement are dominated by the
uncertainties in the absolute mass calibration. To test the effect of
these uncertainties, we changed the normalization of the mass vs.\ proxy
relations by $\pm9\%$ (our estimate of systematic errors in the mass
scale calibration, see \S~\ref{sec:systematics:in:data}). The effect,
shown by the dotted contour in Fig.\,\ref{fig:OmegaM-h}, is to shift the
estimated values of $\sigma_8$ by $\pm 0.02$, just outside the
statistical 68\% CL uncertainties. This range can be considered as a
systematic uncertainty in our $\sigma_8$ determination for a fixed
$\OmegaM$.

Our cluster constraints on $\sigma_8$ are more accurate (for a fixed
$\OmegaM$) than any other method, even including systematic errors
(Fig.\,\ref{fig:OmegaM-h-comp}). It is encouraging that our results are in
very good agreement with recent results from other methods. The
measurements based on lensing sheer surveys, cluster mass function with
$\Mtot$ estimated from galaxy dynamics, and WMAP (5-year results
assuming flat \LCDM{} cosmology) are all within their respective 68\% CL
uncertainties from our best fit. This independently confirms that our
calibration of the cluster mass scale is not strongly biased.
Furthermore, the present systematic errors in the cluster analysis are
smaller than the statistical accuracy provided by WMAP-5 and other
methods. This allows us to effectively use the $\sigma_8$ information in
the dark energy equation of state constraints
(\S\,\ref{sec:wo:clusters+cmb}).

We now move to models where the crucial role is played by the
high-redshift cluster mass function data. The first case to consider is
combined constraints for $\OmegaM$ and $\OmegaL$ in the non-flat \LCDM{}
cosmology. To better demonstrate what role the different components of
the information provided by the cluster mass function play in the
combined constraints, we consider two cases: a) when the full cluster
mass function information is used, and b) when the shape information is
artificially removed thus leaving only the evolutionary information.

\section{Constraints for non-flat \LCDM{} cosmology: $\OmegaM-\OmegaL$}
\label{sec:O-L}

In the first case, for each combination of parameters, we compute the full
likelihood for the low and high-$z$ mass functions and add the HST prior on
the Hubble constant (this is necessary for effective use of the mass
function shape information, see \S\,\ref{sec:fitting} and
\S\,\ref{sec:OmegaMh}). We then marginalize the combined likelihood over
non-essential parameters ($\sigma_8$ and $h$ in this case) keeping the
primordial power spectrum index fixed at the WMAP best-fit value, $n=0.95$.
Removal of the shape information (our second case) is achieved by letting
$n$ vary and marginalizing over it.  This is approximately equivalent to
using a free shape parameter for the CDM power spectra, the approch often
used in earlier cluster studies \citep[e.g.,][]{2001ApJ...561...13B}.
Constraints for both cases were obtained for mass functions estimated using
all our three proxies, $\TX$, $\Mgas$, and $\YX$.

\begin{figure}
\centerline{\includegraphics[width=\columnwidth]{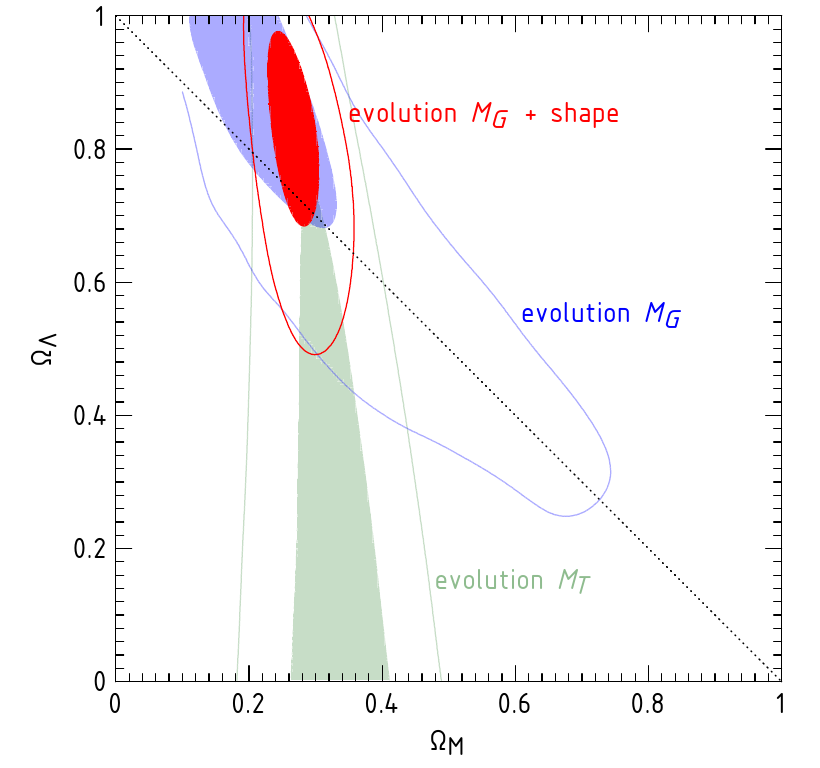}}
\caption{Constraints for non-flat \LCDM{} cosmology from evolution of
  the cluster mass function. The results using only the evolution
  information (change in the number density of clusters between $z=0$
  and $z\approx0.55$) are shown in blue and green from the $\Mgas$ and
  $\TX$-based total mass estimates. The degeneracies in these cases are
  different because these proxies result in very different
  distance-dependence of the estimated masses (see text for details).
  The constraints from the $\YX$-based mass function are between those
  for $\Mgas$ and $\TX$ (Fig.\,\ref{fig:OmegaM-OmegaL:Yx}). Adding the
  shape of the mass function information breaks degeneracies with
  $\OmegaM$, significantly improving constraints from $\Mgas$ and $\YX$
  with little effect on the $\TX$ results.}
\label{fig:OmegaM-OmegaL}
\end{figure}

\begin{figure}
\centerline{\includegraphics[width=\columnwidth]{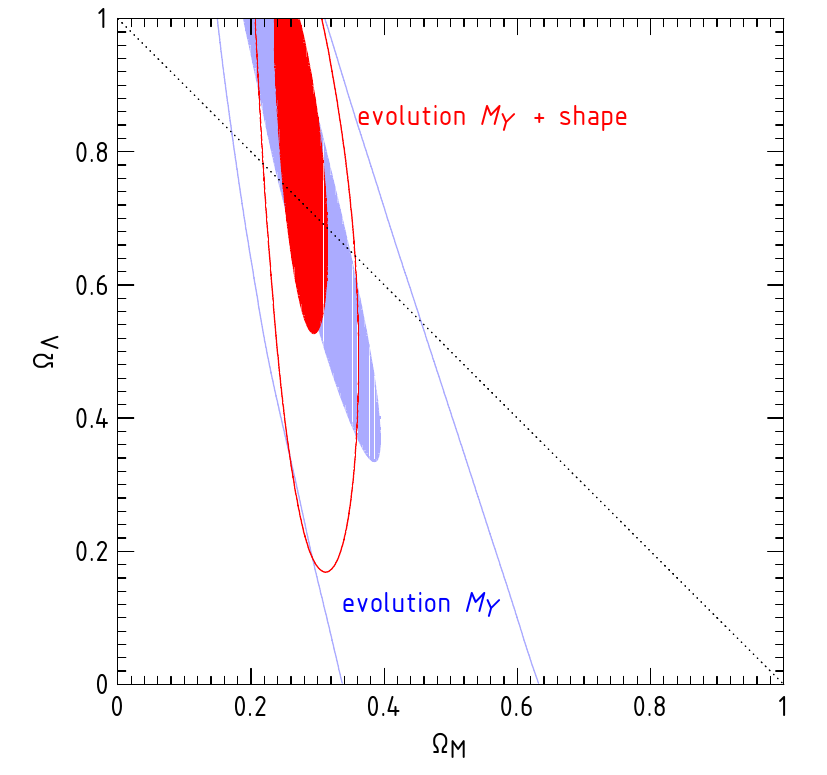}}
\caption{Same as Fig.\,\ref{fig:OmegaM-OmegaL} but for $\YX$-based mass
  estimates.}
\label{fig:OmegaM-OmegaL:Yx}
\end{figure}

The results are presented in Figs.\,\ref{fig:OmegaM-OmegaL}
and~\ref{fig:OmegaM-OmegaL:Yx}. First, we can easily identify the role
of using the mass function shape information (illustrated for the
$\Mgas$ and $\YX$ proxies). Clearly, it mostly breaks the degeneracies
along the $\OmegaM$ axis. The best fit values and statistical
uncertainties for $\OmegaM$ are very close to those derived from the
shape of the local mass function (and nearly identical to those from the
total sample, \S\,\ref{sec:OmegaMh}).
%  This suggests that the shape and
% evolutionary information provided by the cluster mass function are
% separable and thus the corresponding systematic errors can be treated
% independently.  For example, the dominant source of systematic errors on
% $\OmegaMh$ (slope of the $L-M$ relation) will simply extend the
% confidence regions in
% Figs.\,\ref{fig:OmegaM-OmegaL}--\ref{fig:OmegaM-OmegaL:Yx} (by $\pm
% 0.037$ added in quadrature to statistical uncertainties).

For a fixed $\OmegaM$, the observed evolution in the cluster mass
function provides a constraint on $\OmegaL$. Degeneracies in the
$\OmegaM-\OmegaL$ plane provided by different mass proxies applied to
the same set of clusters differ because of the different distance
dependence of the $\Mtot$ estimates via $\TX$, $\Mgas$, and $\YX$ (see
below). Even without the shape information, evolution in the $\YX$ and
$\Mgas$-based mass functions requires $\OmegaL>0$ at the 85\% and 99.7\%
CL, respectively. Including the shape information, we obtain
$\OmegaM=0.28\pm0.04$, $\OmegaL=0.78\pm0.25$ (and $\OmegaL>0$ is
required at the 99\% CL) from the $\YX$-based analysis. The evolution of
the $\Mgas$-based mass function gives $\OmegaM=0.27\pm0.04$,
$\OmegaL=0.83\pm0.15$, and $\OmegaL>0$ at 99.98\% CL. The $\TX$-based
mass function does not strongly constrain $\OmegaL$ but provides an
independent measurement of $\OmegaM$ with almost no degeneracy with
$\OmegaL$: $\OmegaM=0.34\pm0.08$, in good agreement with the mass
function shape results \citep[and also previous measurements based on
evolution of the cluster temperature function,
see][]{2004ApJ...609..603H}. In a flat \LCDM{} model (the one with
$\OmegaM+\OmegaL=1$), the constraint is slightly tighter,
$\OmegaM=0.30\pm0.05$.

Systematic uncertainties of the $\OmegaL$ measurements are dominated by
possible departures of evolution in the $\Mtot$ vs.{} proxy relations.
This issue is discussed in detail below in connection with the dark
energy equation of state constraint (\S\,\ref{sec:w:systematics}); here
we note only that the systematic uncertainties are approximately 50\% of
the purely statistical errorbars on the dark energy parameters
($\OmegaL$, $w$). Therefore, our cluster data provide a clear
independent confirmation for non-zero $\OmegaL$.

%
% Y: deltachi2 = 3.1 - 4.778
% G: 2.4 - 11.0 = 
%
% G+shape:  4.02-18.3029 
% Y+shape:  3.7-8.40894

\paragraph{Comments on the role of geometric information in the cluster
  mass function test} 
Cosmological constraints based on fitting the cluster mass function
generally use not only information from growth of structure but also
that from the distance-redshift relation because derivation of the
high-$z$ mass functions from the data assumes the $d(z)$ and $E(z)$
functions. Quite generally, the estimated mass is a power law function
of these dependencies, $\tilde{M} \propto d(z)^\beta\,
E(z)^{-\varepsilon}$. Different mass proxies have different $\beta$ and
$\varepsilon$, and thus combine the geometric and growth of structure
information in different ways and lead to different degeneracies in the
derived cosmological parameters. We find that strongly
distance-dependent proxies (such as $\Mgas$, see
\citetalias{vikhlinin08}) are intrinsically more powerful in
constraining the dark energy parameters ($\OmegaL$, $w$).  By contrast,
distance-independent proxies such as $\TX$ result in poor sensitivity to
dark energy but instead better constrain $\OmegaM$. This is well
illustrated by the results in Fig.\,\ref{fig:OmegaM-OmegaL}. The $\Mgas$
based estimates for $\Mtot$ result (if we ignore the shape of the mass
function information) in degeneracy approximately along the line
$\OmegaM+\OmegaL=1$. In fact, the evolution of the cluster mass
functions derived from $\Mgas$ can be made broadly consistent with the
$\OmegaM\approx1$, $\OmegaL\approx0$ cosmology if one allows for strong
deviations from the CDM-type initial power spectra
\citep{2006A&A...452...47N}. However, the mass functions estimated from
the temperatures of the same clusters are grossly inconsistent with such
a cosmology, irrespective of the assumptions on the initial power
spectrum ($\OmegaM=1$ is $8.3\sigma$ away from the best fit to the
temperature-based mass function, Fig.\,\ref{fig:OmegaM-OmegaL}).  It is
encouraging that the 68\% CL regions for all three mass proxies overlap
near the ``concordance'' point at $\OmegaM=0.25-0.3$ and
$\OmegaL=0.7-0.75$.

\begin{deluxetable*}{p{3cm}lp{4.0cm}cp{5.0cm}}
  \tablecaption{Parameter constraints from combination of clusters with
    other cosmological datasets\label{tab:par:w}}
  \tablehead{
    \colhead{Parameter} &
    \colhead{Value} &
    \colhead{Dataset} &
    \colhead{Systematic} &
    \colhead{Dominant source}
    \\
    \colhead{}  &
    \colhead{} & 
    \colhead{} &
    \colhead{errors} &
    \colhead{of systematics}
%     \\
%     \colhead{(1)}  &
%     \colhead{(2)} &
%     \colhead{(3)} &
%     \colhead{(4)} &
%     \colhead{(5)}
  }
  \startdata
  \multicolumn{4}{l}{\qquad\emph{Flat ($\OmegaK=0$), constant $w$
      ($w=\wo$)}}\\[1ex]
  $\wo$\dotfill
               & $-1.14\pm0.21$ 
               &\raggedright evol+shape+$h$, \S\,\ref{sec:wo:clusters}
               & $\pm 0.10$, $\pm 0.08$,
               &{\raggedright Evolution of $\Mtot$ vs.\ proxy relations,\\
               evolution in $\LX-\Mtot$, respectively}
\\[\baselineskip]
  $\wo$\dotfill
               & $-1.08\pm0.15$ 
               &\raggedright evol+cmb, \S\,\ref{sec:wo:clusters+cmb}
               & $\pm 0.025$
               &{\raggedright Evolution of $\Mtot$ vs.\ proxy relations}\\
  $\wo$\dotfill
               & $-0.97\pm0.12$ 
               &\raggedright evol+cmb+$\sigma_8$+bao,
               \S~\ref{sec:wo:clusters+cmb}
               & $\pm 0.038$
               &{\raggedright Absolute cluster mass calibration}\\[1\baselineskip]
  $\OmegaX$\dotfill
               & $\phantom{-}0.732\pm0.016$ &\raggedright cmb+bao+sn,
               \S\,\ref{sec:wo:clusters+others}%\tablenotemark{a}
               & \\
  $\wo$\dotfill
               & $-0.995\pm0.067$ 
               &\raggedright cmb+bao+sn,
               \S\,\ref{sec:wo:clusters+others}%\tablenotemark{a}
               & $\pm 0.076$
               &{\raggedright SN systematics}
               \\[1\baselineskip]
  \bfseries\boldmath$\wo$\dotfill
               &\boldmath $-0.991\pm0.045$ 
               &\raggedright cmb+bao+sn+evol+$\sigma_8$,
               \S\,\ref{sec:wo:clusters+others} 
               & $\pm 0.022$, $\pm 0.033$
               &{\raggedright SN systematics, cluster masses}\\
  \bfseries\boldmath$\OmegaX$\dotfill
               &\boldmath $\phantom{-}0.740\pm0.012$
               &\raggedright cmb+bao+sn+evol+$\sigma_8$,
               \S\,\ref{sec:wo:clusters+others} & \\
  \bfseries\boldmath$h$\dotfill
               &\boldmath $\phantom{-}0.715\pm0.012$
               &\raggedright cmb+bao+sn+evol+$\sigma_8$,
               \S\,\ref{sec:wo:clusters+others} &  \\
  \bfseries\boldmath$\sigma_8$\dotfill
               &\boldmath $\phantom{-}0.786\pm0.011$
               &\raggedright cmb+bao+sn+evol+$\sigma_8$,
               \S\,\ref{sec:wo:clusters+others} &  \\
  \hline\\[-1.6ex]
  \multicolumn{4}{l}{\qquad\emph{Flat ($\OmegaK=0$), constant $w$
      ($w=\wo$),  non-zero neutrino mass}}\\[1ex]
  $\wo$\dotfill
               &$-1.02\pm0.055$ 
               &\raggedright $\sigma_8$+cmb$_\nu$+cmb+bao+sn+evol,
               \S\,\ref{sec:mnu}
               & $\pm 0.064$
               &{\raggedright SN systematics.}\\
  $\sum m_\nu$\dotfill
               &$\phantom{-}0.1\pm0.12$~eV,
               &\raggedright $\sigma_8$+cmb$_\nu$+cmb+bao+sn+evol,
               \S\,\ref{sec:mnu}
               & $\pm 0.1$~eV
               &{\raggedright SN systematics, cluster masses}\\
               &$<0.33$~eV (95\% CL)\\
  \hline\\[-1.6ex]
  \multicolumn{4}{l}{\qquad\emph{Flat ($\OmegaK=0$), evolving $w$: $w=w_0+w_a(1-a)$}}\\[1ex]
  $w_a+3.64(1+w_0)$\dotfill
               &$\phantom{-}0.05\pm0.17$
               &\raggedright cmb+bao+sn+evol+$\sigma_8$,
               \S\,\ref{sec:w0-wa} &
               \\
   \hline\\[-1.6ex]
  \multicolumn{4}{l}{\qquad\emph{Constant $w$ ($w=\wo$), non-flat ($\OmegaK\ne0$)}}\\[1ex]
  $w_0$\dotfill
               &$-1.03\pm0.06$
               &\raggedright cmb+bao+sn+evol+$\sigma_8$,
               \S\,\ref{sec:w0-Ok} &
  \enddata
%  \tablenotetext{a}{See also \citet{2008arXiv0803.0547K}.}
  \tablecomments{Codes used in column~3: \emph{evol} --- evolution of
    the cluster mass function; $h$ --- HST prior on Hubble constant;
    \emph{shape} --- shape of the cluster mass function; \emph{cmb} ---
    WMAP-5 distance priors; $\sigma_8$ --- comparison of the
    cluster-derived $\sigma_8$ with the CMB power spectrum normalization
    (reflecting growth of perturbations between $\zCMB$ and $z=0$);
    \emph{bao} --- BAO distance prior; \emph{sn} --- SN~Ia luminosity
    distances; \emph{cmb$_\nu$} --- WMAP-5+BAO+SN constraints on
    neutrino mass (\S\,\ref{sec:mnu}).}
\end{deluxetable*}

\section{Flat Universe with constant dark energy equation of state:
  $\wo-\OmegaX$}
\label{sec:w0}

Next, we study constraints on a constant dark energy equation of state,
$\wo\equiv p_X/\rho_X$, in a spatially flat universe. The analysis using
cluster data only is equivalent to the $\OmegaM-\OmegaL$ case
(\S\,\ref{sec:O-L}). We compute the likelihood for the cluster mass
functions on a grid of parameters: present dark energy density $\OmegaX$
($=1-\OmegaM$), $\wo$, $h$, and $\sigma_8$, then add the HST prior on
the Hubble constant (\S\,\ref{sec:fitting}). Marginalization over
non-essential parameters, $h$ and $\sigma_8$, gives the likelihood as a
function of $\OmegaM$ and $\wo$.  We also obtain the equation of state
constraints combining our cluster data with the three external
cosmological data sets \citep[following the reasoning of][for the choice
of these datasets]{2008arXiv0803.0586D}:

\subsection{External Cosmological Datasets}
\label{sec:other:data}

\paragraph{SN~Ia} We use the distance moduli estimated for the Type~Ia
supernovae from the HST sample of \citet{2007ApJ...659...98R}, SNLS
survey \citep{2006A&A...447...31A}, and ESSENCE survey
\citep{2007ApJ...666..694W}, combined with the nearby supernova sample
\citep[we used a combination of all these samples compiled
by][]{2007ApJ...666..716D}. Calculation of the SN~Ia component of the
likelihood function for the given cosmological model is standard and can
be found in any of the above references.

\paragraph{Baryonic Acoustic Oscillations} Detection of the barynic
acoustic peak in the correlation function for large red galaxies in the
SDSS survey leads to a good measurement of the combination
\begin{equation}
  \left[\frac{d_A(z)^2}{(cz)^2\,H(z)}\right]^{1/3}\!\!\sqrt{\OmegaM H_0^2}\;\;
  \left[\frac{n}{0.98}\right]^{0.35} = 0.469 \pm 0.017
\end{equation}
at $z=0.35$ \citep[``SDSS LRG sample'']{2005ApJ...633..560E}. This prior
mostly constrains $\OmegaM$ but has some sensitivity also to the
dark energy equation of state.

A more recent measurement of the BAO peaks in the combined SDSS and 2dF
survey data is presented in \citet{2007MNRAS.381.1053P} who determine
the BAO distance measure at two redshifts ($z=0.2$ and $z=0.35$) instead
of one in \citet{2005ApJ...633..560E}. These new data are somewhat in
tension ($\sim 2\sigma$) with the SN+WMAP results \citep[see, e.g.,
Fig.\,11 in][]{2007MNRAS.381.1053P}, which may artificially tighten the
constraints when the BAO data are combined with SN~Ia, WMAP, and
clusters. We checked, however, that from the combination of SN~Ia, WMAP,
and SDSS-LRG BAO, we derive the parameter constraints that are
essentially equivalent to those in \citet{2008arXiv0803.0547K}, who used
the \citeauthor{2007MNRAS.381.1053P} priors. Therefore, the choice of
the BAO dataset is unimportant in the combined constraints.

\paragraph{WMAP-5} The likelihood for WMAP 5-year data is computed using a
simplified approach described in \S\,5.4 of \citet{2008arXiv0803.0547K}.
This involves a computation, for a given set of cosmological parameters, of
three CMB parameters --- angular scale of the first acoustic peak, $\ell_A$;
the so called shift parameter, $R$; and the recombination redshift, $z_*$.
The likelihood for the WMAP-5 data is then computed using the covariance
matrix for $\ell_A$, $R$, and $z_*$ provided in
\citeauthor{2008arXiv0803.0547K} This method is almost as accurate as direct
computation of the WMAP likelihood \citep{2007PhRvD..76j3533W} but is much
faster, which allowed us to explore the entire multi-dimensional grid of the
cosmological parameters instead of running Markov chain simulations. One
additional note is that to compute the CMB likelihood, we had to add the
absolute baryon density, $\Omega_bh^2$, to our usual set of cosmological
parameters and then marginalize over it. The reason is that while the
average baryon density has very little impact on the rest of our analysis,
the CMB data are very sensitive to $\Omega_bh^2$, thus any variation of
$h$ must be accompanied by the corresponding variation of $\Omega_b$ without
which the computation of the CMB likelihood would be inadequate.

The method outlined above recovers essentially the entire information
from the location and relative amplitudes of the peaks in the CMB power
spectrum \citep{2007PhRvD..76j3533W}. One additional piece of
information is the absolute normalization of the CMB power spectra,
reflecting the amplitude of density perturbations at the recombination
redshift, $z_*\approx1090$. Contrasted with $\sigma_8$ determined from
our cluster data at $z\approx 0$, it constrains the total growth of
density perturbations between the CMB epoch and the present, and thus is
a powerful additional dark energy constraint.

\paragraph{WMAP-5 plus local $\sigma_8$} The WMAP team provides the
amplitude of the curvature perturbations at the $k=0.02$~Mpc$^{-1}$
scale,
\begin{equation}
  \DeltaR^2 = (2.21\pm0.09)\times10^{-9}
\end{equation}
Section 5.5 in \citet{2008arXiv0803.0547K} gives the prescription of how to
predict this observable for a given set of cosmological parameters and
$\sigma_8$. A useful accurate fitting formula can also be found in
\citet{2004PhRvD..70d3009H}:
\begin{equation}\label{eq:CMB:s8:fit}
  \begin{split}
    \tilde{\Delta}_{\frak R} \approx & \frac{\sigma_8}{1.79\times10^{4}}\,
    \left(\frac{\Omega_b h^2}{0.024}\right)^{1/3}
    \left(\frac{\OmegaM h^2}{0.14}\right)^{-0.563} \\
    & \times\;(7.808\,h)^{(1-n)/2}\,
    \left(\frac{h}{0.72}\right)^{-0.693}
    \frac{0.76}{G_0}
  \end{split}
\end{equation}
(we adjusted numerical coefficients to take into account that the
\citeauthor{2004PhRvD..70d3009H} approximation uses the CMB amplitude at
$k=0.05$~Mpc$^{-1}$ while the WMAP-5 results are reported for
$k=0.02$~Mpc$^{-1}$). In this equation, $G_0$ is the perturbation growth
factor between the CMB redshift and the present, normalized to the
growth function in the matter-dominated universe: $G(z) \equiv (1+z)\,
\delta(z)/\delta(z_{\rm CMB})$. This fitting formula helps to understand
the nature of the $\sigma_8$ vs.{} CMB amplitude constraint. The
relation between $\sigma_8$ and $\DeltaR$ depends on the absolute matter
and baryon densities, $\OmegaM h^2$ and $\Omega_b h^2$ (well-measured by
the CMB data alone), and on the total growth factor, $G_0$, and the
absolute value of the Hubble constant, $h$. Both of these quantities
provide powerful constraints on any parametrization of the dark energy
equation of state \citep{2005ASPC..339..215H}, and their combination
does so as well.

Inclusion of this information in the total likelihood is
straightforward. Given the usual set of cosmological parameters
($\OmegaX$, $\wo$, $h$) plus $\sigma_8$, one computes
\begin{equation}
  \chi^2_{\rm CMBnorm} = (\tilde{\Delta}_{\frak R}^2\times10^9 -
  2.21)^2/0.09^2,
\end{equation}
where $\tilde{\Delta}_{\frak R}$ can be obtained either from
eq.[\ref{eq:CMB:s8:fit}] or as described in \citet{2008arXiv0803.0547K}.
The $\chi^2_{\rm CMBnorm}$ component is then added to the cluster
$\chi^2$ and the sum marginalized over $\sigma_8$.

\subsection{$\wo$ from Cluster Data Only}
\label{sec:wo:clusters}

\begin{figure}
\centerline{\includegraphics[width=\columnwidth]{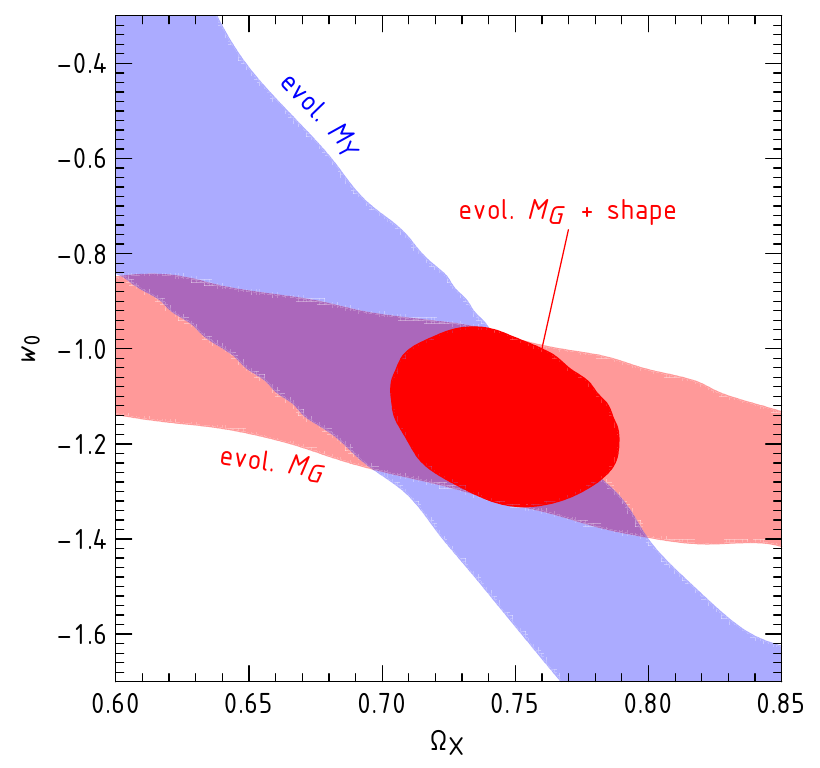}}
\caption{Constraints on the present dark energy density $\OmegaX$ and
  constant equation of state parameter $\wo$ derived from cluster mass
  function evolution in a spatially flat Universe. The results for
  $\Mgas$ and $\YX$-based total mass estimates are shown in red and
  blue, respectively. The inner solid red region shows the effect of
  adding the mass function shape information (\S\ref{sec:OmegaMh}) to
  the evolution of the $\Mgas$-based mass function.}
\label{fig:w0:clusters}
\end{figure}

Constraints on the present dark energy density $\OmegaX$ and constant
equation of state are presented in Fig.\,\ref{fig:w0:clusters}. For
comparison, we show separately the results derived only from evolution
of the $\Mgas$ and $\YX$-based mass functions, and the effect of
including the mass function shape information (\S\,\ref{sec:O-L}
describes the procedure for removing shape information from the cluster
likelihood function). We do not consider here the $\TX$ based mass
estimates because they provide little sensitivity to the dark energy
parameters (\S\,\ref{sec:O-L}). Just like in the $\OmegaM-\OmegaL$ case,
evolution of the $\Mgas$ and $\YX$-based mass functions constrains
different combinations of $w_0$ and $\OmegaX$.  The width of the
confidence regions across the degeneracy direction is similar but the
gas-based results are less inclined giving a little more sensitivity to
$w_0$ for a fixed dark energy density~--- $\Delta \wo = \pm 0.17$ from
the $\Mgas$-based functions and $\Delta \wo = \pm 0.26$ from $\YX$.

\begin{figure}
\centerline{\includegraphics[width=\columnwidth]{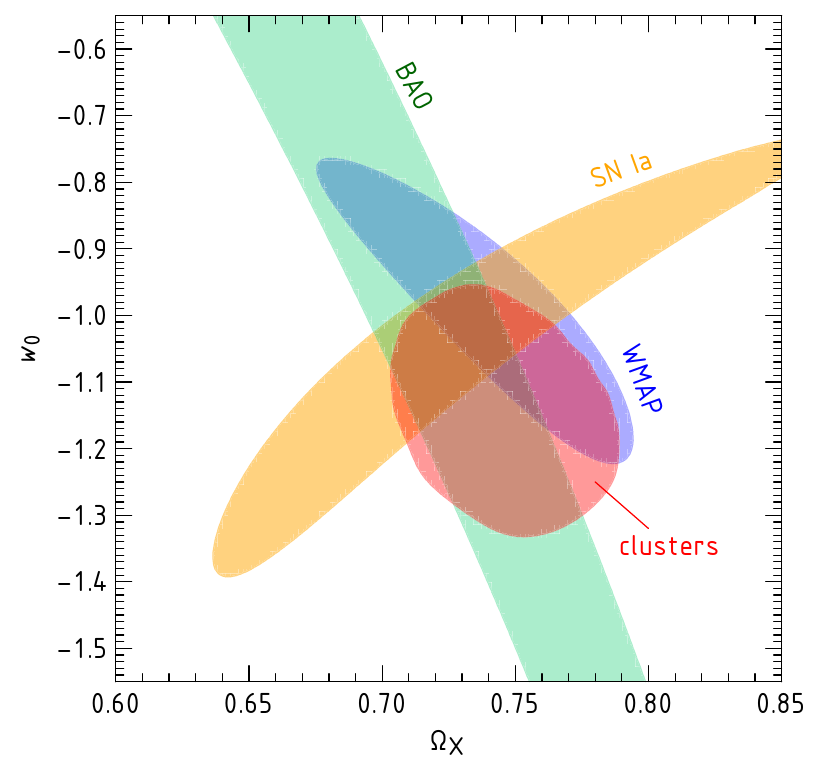}}
\caption{Comparison of the dark energy constraints from X-ray clusters
  and from other individual methods (supernovae, baryonic acoustic
  oscillations, and WMAP).}
\label{fig:w0:comparison}
\end{figure}

Adding the mass function information combined with the HST prior on $h$
breaks the degeneracy along the $\OmegaX$ direction. For example, the
ellipse in Fig.\,\ref{fig:w0:clusters} shows the 68\% CL region from fitting
both the evolution and shape of the $\Mgas$-based mass function.  The
one-parameter confidence intervals in this case are $\OmegaX=0.75\pm0.04$
and $\wo=-1.14\pm0.21$. These results compare favorably with those from
other individual methods --- supernovae, BAO, WMAP
(Fig.\,\ref{fig:w0:comparison}), although the supernovae and CMB data
provide tighter constraints on $\wo$ for a fixed $\OmegaX$. The real
strength of the cluster data is, however, when they are combined with the
CMB and other cosmological datasets. The combined constraints are very
similar for the $\Mgas$ and $\YX$-based cluster mass functions, and
therefore we discuss only the former hereafter.

\subsection{$\wo$ from the Combination of Clusters with Other Data}
\label{sec:wo:clusters+cmb}
\label{sec:wo:clusters+others}

\begin{figure}
\centerline{\includegraphics[width=\columnwidth]{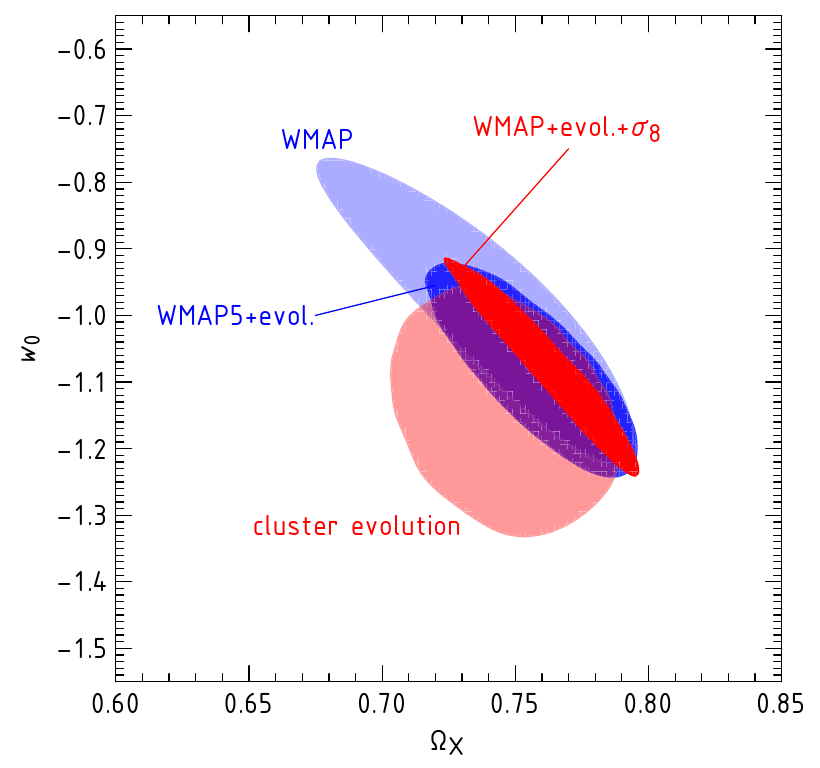}}
\caption{Dark energy constraints in a flat universe from the combination
  of the CMB and cluster data (dark blue region). Adding the $\sigma_8$
  vs.\ CMB normalization information significantly improves constraints
  on $\wo$ for a fixed $\OmegaX$ (inner red region).}
\label{fig:w0:clusters+cmb}
\end{figure}

First, we consider a combination of the cluster data with the WMAP distance
priors \citep[see \S\,5.4 in][]{2008arXiv0803.0547K}. Cluster data bring
information on growth of density perturbations and normalized distances in
the $z\simeq0.0-0.9$ interval, and --- weakly --- on the $\OmegaMh$
parameter. Adding this information reduces the WMAP-only uncertainties on
$\wo$ and $\OmegaX$ approximately by a factor of 2 (dark blue region in
Fig.\,\ref{fig:w0:clusters+cmb}): $\wo = -1.08\pm0.15$,
$\OmegaX=0.76\pm0.04$.

A much more significant improvement of the constraints arises from the
$\sigma_8$ determination from low-redshift clusters (dark red region in
Fig.\,\ref{fig:w0:clusters+cmb}). Comparison of the local determination
of $\sigma_8$ with the CMB normalization mostly provides a measurement
of the total perturbation growth factor between $\zCMB$ and the present.
This depends more sensitively on $\wo$ than the evolution of the cluster
mass function because of, first, larger redshift leverage, and second,
because the perturbation amplitude at high $z$ is measured more
accurately by CMB than by 37~clusters from the 400d survey.

Is it appropriate to use the $\sigma_8$ vs.\ CMB normalization
information in the dark energy constraints or does it require
unreasonable interpolation of the dark energy parametrization to high
redshifts? We note in this regard that for any combination of the
cosmological parameters in the vicinity of the ``concordance'' model,
$\wo\simeq-1$, $\OmegaX=0.25-0.3$, the Universe becomes matter-dominated
and enters the deceleration stage by $z\sim 1.5-2$; the growth of
perturbations is basically fixed after that at $G(z)=1$. In other words,
the CMB data can be used to safely predict the amplitude of density
perturbations at $z=1.5-2$ almost independently of the exact dark energy
properties. As long as it is appropriate to use a particular dark energy
parametrization in the $z=0-2$ interval, it is therefore appropriate to
use the same model for the joint clusters$+$WMAP fit.

By itself, adding the $\sigma_8$ information does not significantly
improve the $\wo$ and $\OmegaX$ constraints (the total extent of the
$1\sigma$ confidence regions is similar to the WMAP+evolution case), but
the confidence region becomes much more degenerate with $\OmegaX$ (see
inner red region in Fig.\,\ref{fig:w0:clusters+cmb}), which increases
the potential for improvement when we combine these results with other
cosmological datasets, BAO and supernovae.
% 
% Adding the BAO prior to the CMB+clusters data gives $\wo=-0.97\pm0.12$,
% and $\OmegaX=0.73\pm0.025$. The crucial improvement occurs, however,
% when we combine the CMB+clusters+BAO results with the supernovae data,
% because the SN~Ia degeneracy in the $\OmegaX-\wo$ plane is more
% orthogonal to those for other datasets than the WMAP, BAO, and cluster
% results to each other (Fig.\,\ref{fig:w0:comparison}). 

\begin{figure}
\centerline{\includegraphics[width=\columnwidth]{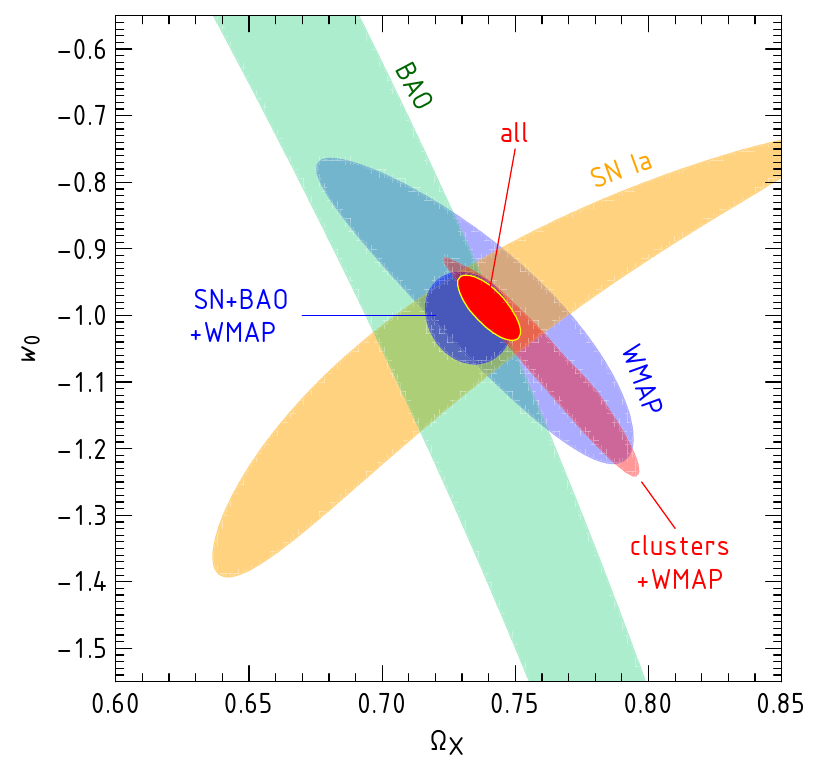}}
\caption{Dark energy constraints in flat universe from combination of
  all cosmological datasets. We find $\wo=-0.991\pm0.045$ ($\pm 0.04$
  systematic) and $\OmegaX=0.740\pm0.012$, see Table~\ref{tab:par:w} and
  \S\,\ref{sec:wo:clusters+others}.}
\label{fig:w0:all}
\end{figure}

The combined constraints from all four cosmological datasets are shown
in Fig.\,\ref{fig:w0:all} (inner dark red region). The 68\%
one-parameter confidence intervals are $\OmegaX=0.740\pm0.012$ and
$\wo=-0.991\pm0.045$.  The importance of adding information from our
cluster samples is illustrated by a factor of $\sim 1.5$ reduction of
the measurement uncertainties with respect to the WMAP+SN+BAO data
alone: we obtain $\wo=-0.995\pm0.067$ without clusters \citep[dark blue
region in Fig.\,\ref{fig:w0:all}; these results are essentially
identical to those reported in][]{2008arXiv0803.0547K}.  Perhaps more
importantly, including the cluster data also reduces systematic
uncertainties by a similar amount (\S\,\ref{sec:w:systematics}).

The best-fit values of the Hubble constant and $\sigma_8$ from the
combination of all datasets are $h=0.715\pm0.012$ and
$\sigma_8=0.786\pm0.011$. These values are within $68\%$ confidence
intervals of their determination by direct measurements (HST Key Project
results for $h$ and fitting the low-$z$ cluster mass function for
$\sigma_8$). The best-fit combination of the dark energy parameters is
also within the $1\sigma$ confidence regions for each individual dataset
included in the constraints (Fig.\,\ref{fig:w0:all}). Therefore, the
best-fit cosmological model is a \emph{good fit} to the data. In particular,
Fig.\,17 from \citetalias{vikhlinin08} shows that the mass function models
computed in the \LCDM{} cosmology ($w_0=-1$) provide a very good description
of the data.

\subsection{Systematic Uncertainties in the $\wo$ Measurements}
\label{sec:w:systematics}

We estimate the effect of known sources of systematics on the cosmological
constraints by varying the corresponding individual sets of data or internal
relations (e.g., evolution in $\LX-\Mtot$ entering the survey volume
computations) within the estimated $1\sigma$ interval. We assume,
optimistically, that the current WMAP and BAO data are free from significant
systematics (i.e., that they are smaller than statistical uncertainties),
and consider systematic errors only in the SN~Ia and cluster datasets. In
most cases, a single source clearly dominates the systematic error budget
for a particular measurement, so we report on only those dominant sources.

The largest known source of systematic error in the SN~Ia analysis is
the correction for extinction in host galaxies and uncertainties in
intrinsic colors of SN~Ia \citep[e.g.,][]{2008arXiv0803.0982F}. As a
measure of systematic uncertainty in the combined SN sample we use
$\pm0.13$ in $\wo$ for fixed $\OmegaX$, quoted by
\citet{2007ApJ...666..694W}. We implement these errors by computing the
SN likelihood in our experiments for $(\OmegaX,\wo+0.13)$ and
$(\OmegaX,\wo-0.13)$ instead of $(\OmegaX,\wo)$.

\subsubsection{Main Sources of Cluster Sustematics}

The largest sources of systematic errors in the cluster analysis are
those in the normalization of the $\Mtot$ vs.\ proxy relations. They can
be separated into two almost independent components: 1) how accurately
is the \emph{absolute} cluster mass scale established by X-ray
hydrostatic $\Mtot$ estimates in the low-redshift clusters, and 2) how
accurately can we predict evolution in the $\Mtot$ vs.\ proxy relations,
i.e., the \emph{relative} mass scale between low and high redshift
clusters. The first component mainly affects the $\sigma_8$ measurements
and associated dark energy constraints, while the second component
affects the results derived from using only evolution in the cluster
mass function (those in Fig.\,\ref{fig:w0:clusters}). Our estimates of
the $\Mtot$ systematics are discussed extensively in
\citetalias{vikhlinin08}. For the absolute mass scale ($\Mtot$ for fixed
$\YX$, $\TX$, or $\Mgas$) at $z\approx0$, we estimate $\Delta M_{\rm
  sys}/M\lesssim 9\%$ mainly from comparison of the X-ray and weak
lensing mass estimates in representative samples. This source of error
is implemented by changing the normalization of the $\Mtot$ vs.\ $\YX$,
$\Mgas$, or $\TX$ relations at $z=0$ by $\pm9\%$. For uncertainties in
the evolution of the $\Mtot$ vs.\ proxy relations, we estimate $\Delta
M/M \approx 5\%$ at $z=0.5$, mainly from comparison of the prediction of
different models describing observed small deviations of the cluster
scaling relations from self-similar predictions, and from the magnitude
of these deviations and corresponding corrections we apply to the data.
These uncertainties are implemented by multiplying the standard scaling
relations by factors of $(1+z)^{\pm0.12}$.

Comparable to the evolution in the $\Mtot$ vs.\ proxy relation are
measurement uncertainties in the evolution factor for the $\LX-\Mtot$
relation. We do not use $\LX$ to estimate the cluster masses, but the
relation is required to compute the survey volume for the high-$z$
sample.  The resulting volume uncertainty depends on the mass scale, and
can become comparable to the Poisson error for the comoving cluster
number density \citepalias[see \S\,5.1.3 in][]{vikhlinin08}. We tested
how this influences the cosmological fit by varying the parameters of
the $\LX-\Mtot$ relation within their measurement errors around the best
fit [the evolution of $\LX$ for fixed $\Mtot$ in our model is
parametrized as $E(z)^\gamma$ and $\gamma$ is measured to $\pm0.33$, see
\S\,5.1.3 in \citetalias{vikhlinin08}].

Other sources of systematics in the cluster analysis \citepalias[summarized
in][]{vikhlinin08} are negligible compared to those outlined above. We
verified also that uncertainties in the intrinsic scatter in the
$\Mtot$-proxy relations are not important. The main reason is that in the
dark energy constraints, we use high-quality mass proxies ($\YX$ and
$\Mgas$), which should provide mass estimates with small, 7--10\% scatter.
Variations of this scatter by up to $\pm 50\%$ with respect to the nominal
values do not significantly change the best fit cosmological parameters.
This conclusion is seemingly different from \citet{2005PhRvD..72d3006L}
because in that paper, they consider proxies with larger scatter (the effect
on the cosmological parameter constraints is proportional to scatter
squared), and also they assumed that the normalizations in the $\Mtot$ vs.\
proxy relation are obtained from self-calibration while we use direct mass
measurements for a well-observed subsample.

The variations of the best-fit parameters due to the systematics
discussed above are reported in Table~\ref{tab:par:w} along with the
dominant source of error for each combination of cosmological datasets.
For example, variations in the evolution of the $\Mtot-\Mgas$ and
$\Mtot-\YX$ relations affect the best fit to the cluster data only by
$\Delta\wo=\pm0.1$, while statistical uncertainties are $\pm0.2$ to
$\pm0.3$ for fixed $\OmegaX$ (\S\,\ref{sec:wo:clusters}); unless the
systematics in this case are a factor of two larger than our estimates,
they are unimportant.
\looseness-1

\subsubsection{Systematics in the Combined Constraints}

The most interesting case to consider is reduction in the systematic
errors from combining both SN and cluster data with the WMAP and BAO
priors. In the SN+CMB+BAO case, the supernovae systematics cause
variations in the best-fit $\wo$ by $\pm0.076$ (reduced from $\pm0.13$
for the SN-only case mainly by including WMAP priors). Cluster
systematics affects the $\wo$ constraints from the clusters+WMAP+BAO
combination by $\pm0.04$ (dominated by the $\pm9\%$ uncertainties in the
absolute mass scale).  The influence of both sources of error is
significantly reduced in the combined constraints.  We find that the
best fit $\wo$ from SN+clusters+WMAP+BAO is affected by $\pm 0.022$ by
SN systematics, and by $\pm 0.033$ by cluster systematics. The total
systematic error in the combined constraint is thus $\Delta\wo=\pm0.04$,
almost a factor of 2 reduction from $\pm0.076$ achievable without
clusters.

We also note that if we significantly underestimate the cluster
systematics, the most likely direction is that the cluster total masses
are underestimated\footnote{The X-ray hydrostatic analysis includes only
  the gas thermal pressure and assumes that the cluster gas body is
  close to being spherically symmetric. The presence of additional
  components in the pressure, clumpiness and turbulent motions in the
  gas all lead to underestimation of $\Mtot$ derived from X-ray data.
  Probably the only possibility for \emph{overestimation} of $\Mtot$ in
  the X-ray analysis is a gross miscalibration of the \emph{Chandra}
  spectral response, for which strong experimental limits are available.
  \label{fn:mass:unc}}. If cluster $\Mtot$ are revised high, this would lead
to an increase in the derived $\sigma_8$, and decrease in $\wo$ when cluster
data are combined with the CMB priors. Dark energy models predicting the
equation of state parameter significantly above $\wo=-1$ will be even less
consistent with observations in this case.

\subsubsection{Prospects for Futher Reduction of Systematic Errors}

It is reassuring that all sources of systematic errors we considered
affect the dark energy equation of state constraints within the
statistical measurement errors. This implies that while systematic
errors are important, they do not yet dominate the current error budget.
The situation will reverse in the future as the datasets expand. More
effort will be needed then to reduce the systematics still further. We
briefly outline the prospects for reducing the cluster-related
systematics. Some of this will happen automatically as the high-$z$
surveys become deeper and cover a larger area.  For example, the $V(M)$
uncertainties for our range of redshifts can be eliminated simply by
decreasing the flux threshold by a factor of $\sim 4$ compared to the
\400d{} limit, making the sample volume-limited; such an extension will
provide also a more accurate measurement of the $\LX-\Mtot$ relation.
The absolute calibration of $\Mtot$ in low-$z$ clusters can be improved
by constraining sources of non-thermal pressure (e.g., if turbulence is
of any importance for the $\Mtot$ estimates, it is easily detectable
with an X-ray microcalorimeter), or through stacked weak lensing
analysis (e.g., measuring average lensing shear profiles for a large set
of clusters with the same $\YX$). To improve limits on non-standard
evolution in the $\Mtot$ vs.\ proxy relations, we cannot use direct mass
measurements of the high-$z$ objects because they will be degenerate
with the assumed distance-redshift relation. Instead, we should improve
reliability of numerical models for cluster evolution. The biggest
uncertainties in these models at present are related to the processes of
gas cooling and star formation, and also to energy feedback from the
central AGN. The strategy for future progress can be based on the fact
that these processes most strongly affect cluster cores, which we do not
use for the mass estimates. We can, therefore, use the data from the
central regions to bracket a likely range of uncertainty in the model
predictions for the cluster outer regions, where we derive the $\Mtot$
proxies. However, even with the current estimated uncertainties, the
samples can grow by a factor of $\sim 4$ before the systematics start to
dominate. Ultimately, as the cluster surveys detect $\sim 10^4$ clusters
with accurately measured X-ray parameters, the so-called
self-calibration techniques
\citep{2004ApJ...613...41M,2004PhRvD..70d3504L} can be employed to
further constrain the evolution in the $\Mtot$ vs.\ proxy relations.

% \begin{verbatim}
%  OmegaM=0.26
% GAS:
%             evol evol+shape evol+cmb  clus+cmb clus+cmb+bao clus+cmb+BAO+SN evol+shape+SN
% default : -1.154  -1.154    -1.025     -0.984        -0.990         -0.991
% m+9\%   : -1.177  -1.177    -1.031     -1.023        -1.028         -1.024
% evol+5\%: -1.053  -1.053    -1.000     -0.994        -0.999         -1.000
% gamma-0.33 -1.060
% sum_m_nu=0.17: --   ----     -----     -1.096        -1.100         -1.080
% 
% YX:
% default : -1.237  -1.237    -0.978   -0.992   -0.982        -0.987         -1.015
% m+9\%   : -1.290  -1.290    -1.012   -1.026   -1.017        -1.014         -1.019
% evol+5\%: -1.027  -1.027    -0.990   -1.000   -0.990        -0.993         -1.004
% 
% SN:           SN       SN+cmb    SN+cmb+bao     SN+cmb+bao+clus
% default:     -1.00     -0.995     -1.013        -0.991
% sn+0.13:     -1.13     -1.079     -1.089        -1.013
% 
% \end{verbatim}

\begin{figure}
\centerline{\includegraphics[width=\columnwidth]{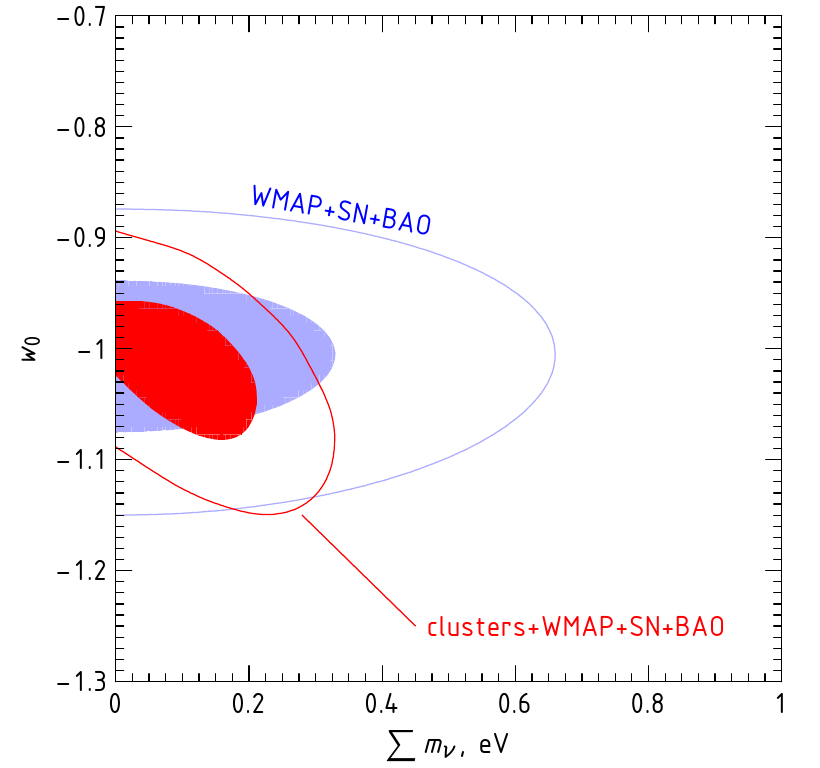}}
\caption{Equation of states from WMAP, BAO, SN~Ia, and clusters in the
  case of non-zero neutrino mass.}
\label{fig:w0-mnu}
\end{figure}

\subsection{Effects of Non-Zero Neutrino Mass}
\label{sec:mnu}

If light neutrinos have masses in the range of a few~$0.1$~eV, they
become non-relativistic between $\zCMB$ and $z=0$, and this transition
produces distortions in the matter perturbations power spectrum relative
to prediction of the pure CDM+baryons model. Using approximations of the
transfer function from \citet{1999ApJ...511....5E}, it is easy to verify
that the effect is approximately proportional to the total mass of
neutrinos (more exactly, to $\sum m_\nu/\OmegaM$), and the rms
fluctuations at cluster scales today are suppressed by approximately
$20\%$ if $\sum m_\nu=0.5$~eV and $\OmegaM=0.26$.  This effect is far
outside the measurement uncertainties in $\sigma_8$ from clusters (we
quote systematic errors of 3\% from uncertainties in the $\Mtot$
calibration and statistical uncertainties are even smaller, see
Table~\ref{tab:par:clusters}).  Therefore, neutrino masses in this range
a) may affect the dark energy constraints when cluster data are combined
with WMAP (because they will effectively change the relation between
$\sigma_8$ and the CMB normalization, eq.\,\ref{eq:CMB:s8:fit}), and b)
can be tightly constrained by our cluster data.

To test the effect of neutrinos, we ran an additional set of models in
which the total neutrino mass was allowed to vary between 0 and 1~eV.
For simplicity we assumed that there are 3 neutrino species with the
same mass, but the final results are not very sensitive to this
assumption.  The only component of our procedure which is significantly
affected by non-zero neutrino mass is contrasting the cluster-derived
$\sigma_8$ with the WMAP normalization of the CMB power spectrum. We can
no longer rely on eq.(\ref{eq:CMB:s8:fit}) and should instead use the
full procedure described in \S\,5.5 of \citet{2008arXiv0803.0547K}.
Otherwise, the analysis is equivalent to the $\sum m_\nu=0$ case. The
likelihood for all cosmological datasets was computed on our usual grid
plus $\sum m_\nu$ as an additional free parameter, and then marginalized
over $\OmegaX$, $h$, and $\sigma_8$.  Finally, we took into account that
a combination of WMAP, BAO, and SN data provides some sensitivity to
neutrino mass through the so-called early integrated Sachs-Wolfe effect
\citep[see discussion in \S\,6.1.3 of][and references
therein]{2008arXiv0803.0547K}. From this analysis,
\citeauthor{2008arXiv0803.0547K} derive a 95\% upper limit of $\sum
m_\nu<0.66$~eV. Since our procedure of using WMAP priors
(\S\,\ref{sec:other:data}) ignores this additional information, we
included it approximately by adding a Gaussian prior $\sum
m_\nu=0\pm0.33$~eV to the final marginalized likelihood.

The derived constraints on $\sum m_\nu$ and $\wo$ are shown in
Fig.\,\ref{fig:w0-mnu}. As expected, when the $\sigma_8$ vs.\ CMB
normalization constraint is added, there is a degeneracy between the
best-fit $\wo$ and total neutrino mass. If we were using only clusters
and WMAP, the degeneracy would approximately follow the line $\wo +1 =
-0.4\sum m_\nu$ and would extend to $\sum m_\nu\approx 1.3$~eV
\citep[the WMAP-only bound on neutrino mass,][]{2008arXiv0803.0586D}.
This degeneracy is broken, however, when we add the BAO and SN
information: low values of $\wo$ required by clusters+CMB for high
values of neutrino mass are inconsistent with these two datasets.
Therefore, a combination of all four datasets can be used to constrain
both $\wo$ and neutrino mass. The best fit value is $\sum
m_\nu=0.10\pm0.12$~eV, with a 95\% CL upper limit of $\sum
m_\nu<0.33$~eV. This limit is significantly tighter than that achievable
without clusters ($<0.66$~eV at 95\% CL). The constraint on $\wo$
degrades somewhat compared to the $m_\nu=0$ case: $\wo=-1.02\pm0.055$
(compared to $\pm0.045$ for $m_\nu=0$), but is still better than
$\pm0.067$ without clusters (see Table~\ref{tab:par:w}). To conclude,
adding the cluster information allows us to set tight limits on the
neutrino mass while still improving the $\wo$ measurements with respect
to the SN+WMAP+BAO case.

Our constraints on neutrino mass are still weaker than the published
results from Ly-$\alpha$ forest data, $\sum m_\nu<0.17$~eV
\citep{2006JCAP...10..014S}. Both the cluster and Ly-$\alpha$ based
constraints use the same effect --- suppression of the power spectrum at
small scales by neutrinos, --- but they have completely different
systematics. The main unknown in the Ly-$\alpha$ analysis is the thermal
state of the low-density IGM, usually estimated from numerical
simulations; it has been suggested that the thermal state may be more
complex than assumed in previous work thus significantly weakening the
$m_\nu$ bounds \citep{2008MNRAS.tmp..423B}.  For clusters, the main
uncertainty is the absolute mass calibration for low-redshift objects
which affects the measurement of $\sigma_8$ (\S\,\ref{sec:s8}). The 9\%
systematic uncertainties on $\Delta M/M$ that we quote would translate
into approximately $\pm0.075$~eV for $\sum m_\nu$, negligible compared
to the current statistical uncertainties. We note that if the X-ray
cluster mass measurements are wrong by more than 9\%, it is almost
certainly in the sense that they are underestimated (see
footnote~\ref{fn:mass:unc} on page~\pageref{fn:mass:unc}); the true
value of $\sigma_8$ will then be higher than our measurement and the
bound on neutrino mass will be even tighter. Therefore, our 95\% CL
bound of $\sum m_\nu<0.33$~eV can be considered as a \emph{conservative}
upper limit.

\section{More general dark energy models}
\label{sec:w:general}

Finally, we demonstrate how our cluster data improves parameter
constraints for more general dark energy models. We consider two cases
--- evolving equation of state, $w=w(z)$, and constant equation of state
in a non-flat universe. The results are presented less completely than
for the case of constant $w$ in a flat universe. We also do not discuss
systematic uncertainties separately for these cases; we checked that the
importance of different sources of systematics and their fraction of
statistical uncertainties is approximately the same as reported in
\S\,\ref{sec:w:systematics} for the constant $w$, flat universe case.

\subsection{$\wz$ in Flat Universe}
\label{sec:w0-wa}

We consider an often used parametrization of the equation of state
evolution in which $w$ changes linearly with the expansion factor, $w(a)
= w_0 + w_a(1-a)$, or equivalently, $w(z) = w_0 + w_a z/(1+z)$. We do
not consider more complex parametrizations because constraints on the
evolution term are still weak, and because neither parametrization has a
clear physical motivation.

\begin{figure}
\centerline{\includegraphics[width=\columnwidth]{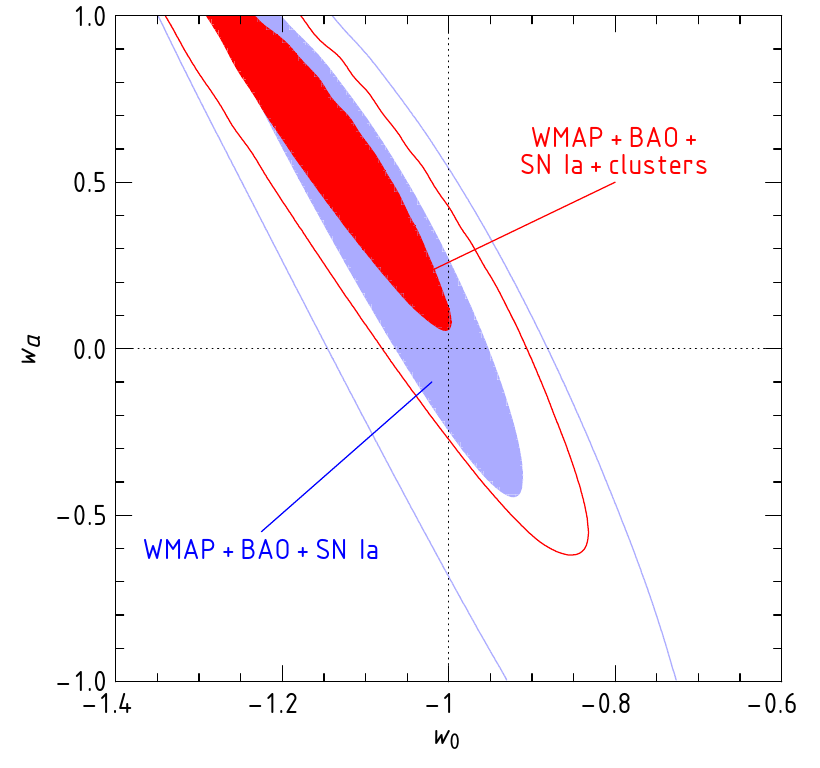}}
\caption{Constrains on evolving equation of state, $w(z) = w_0 + w_a
  z/(1+z)$, in flat universe.}
\label{fig:w0-wa}
\end{figure}

The likelihood function is computed on the $\OmegaM$, $w_0$, $w_a$, $h$,
$\sigma_8$ grid and then marginalized over $\OmegaM$, $h$, and $\sigma_8$,
leading to constraints in the $\wo-\wa$ plane shown in Fig.~\ref{fig:w0-wa}.
Constraints on $w_a$ are weak with or without clusters. For example, the
model with $w_0=-1.2$ and $w_a=1$ (leading to $w=-0.7$ by $z=1$) is
perfectly consistent with the data. However, clusters make the confidence
region substantially narrower (improve $w_a$ constraints for a fixed $w_0$).
A cosmological constant model $(w_0=-1,w_a=0)$ is still consistent with the
data.

Finally, we note that in either case, the degeneracy between $w_0$ and
$w_a$ is almost linear, $w_a = A + B\,w_0$. For such degeneracies,
constraints on constant $w$ are equivalent to those for evolving $w$ at
the pivot redshift, $a_p = (1+z_p)^{-1} = 1+1/B$
\citep{2004PhRvD..70d3009H}. From the slopes of degeneracies in
Fig.\,\ref{fig:w0-wa}, we find $z_p\approx0.29$ without clusters and
$z_p\approx0.38$ when cluster information is included. Therefore, our
combined constraints on constant $w$ (\S\,\ref{sec:w0}) can also be
interpreted as those for evolving $w$ at this pivot redshift.

\subsection{$\wo$ in Non-Flat Universe}
\label{sec:w0-Ok}

\begin{figure}
\centerline{\includegraphics[width=\columnwidth]{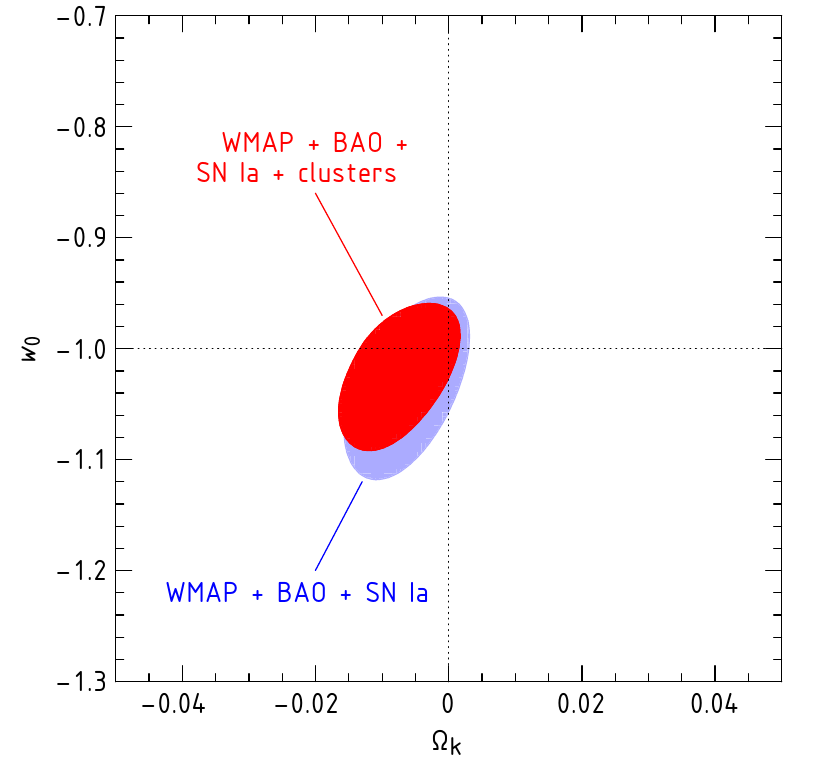}}
\caption{Equation of state constrains from WMAP, BAO, SN~Ia, and
  clusters in the case of non-flat universe. We find $\wo=-1.03\pm0.06$
  and $\Omega_k=-0.008\pm0.009$ with all the data combined.}
\label{fig:w0-Ok}
\end{figure}

The final case we consider is constant $w$ in a non-flat universe. The
cosmological grid in this case is $(\OmegaK, \OmegaM, w_0, h, \sigma_8)$
with the requirement that the dark energy density is
$\OmegaX=1-\OmegaM-\OmegaK$. The likelihood is marginalized over
$\OmegaM$, $h$, $\sigma_8$, and the constraints on $\OmegaK$ and $\wo$
are shown in Fig.\,\ref{fig:w0-Ok}. Including clusters does not
noticeably improve the measurement of $\OmegaK$; by far the most
significant contribution to the $\OmegaK$ constraint is from combination
of WMAP and BAO data \citep{2008arXiv0803.0547K}. However, clusters do
substantially improve the equation of state measurement:
$\wo=-1.03\pm0.06$ to be compared with $\pm0.085$ without clusters.  A
flat \LCDM\ model ($\OmegaK=0$, $w_0=-1$) is still consistent with the
data within 68\% CL.

\section{Summary and conclusions}

We presented constraints on the cosmological parameters from a new
measurement on the galaxy cluster mass function in the redshift range
$z=0-0.9$. All major sources of information contained in the cluster mass
function --- its overall normalization and slope at $z=0$, and evolution at
high redshifts --- are determined with our new data with a higher
statistical accuracy and smaller systematic errors than before. This leads
to much improved and more reliable constraints on the cosmological
parameters. 

From the normalization of the mass function estimated at low redshifts,
we derive the $\sigma_8$ parameter degenerate with $\OmegaM$:
$\sigma_8(\OmegaM/0.25)^{0.47} =0.813\pm 0.013$ (stat) $\pm0.024$ (sys).
The slope of the low-$z$ mass function is a measure of $\OmegaMh$:
$\OmegaMh=0.184\pm0.037$; combined with the HST prior on $h$, this is an
independent measurement of $\OmegaM=0.255\pm0.043$. The matter density
can be independently measured with our cluster data using evolution of
the temperature function, yielding consistent results,
$\OmegaM=0.30\pm0.05$ in a flat \LCDM{} model and $0.34\pm0.08$ in a
general cosmology.

Evolution of the mass functions between $z=0$ and $0.5$ (median redshift
for our high-$z$ sample) constrains $\OmegaL=0.83\pm0.15$ in non-flat
\LCDM{} cosmology, or the dark energy equation of state parameter,
$\wo=-1.14\pm0.21$, in a spatially flat Universe. Inclusion of the
information provided by our cluster data also significantly improves the
equation of state constraints obtained from combination of multiple
cosmological datasets. For example, by combining the 5-year WMAP, most
recent supernovae measurements, and detection of baryonic acoustic
oscillations in the SDSS with our cluster data, we obtain
$\wo=-0.991\pm0.045$ (stat) $\pm0.040$ (sys); both the statistical and
systematic errors in the combined constraint are a factor of $1.5-2$
smaller than those without clusters. Including cluster information also
improves results for an evolving equation of state parameter and for
constant $w$ in a non-flat universe. A spatially flat \LCDM{} model is
within the 68\% CL interval from the best fit in all cases that we
tested.

A good agreement between the geometric and growth of structure-based
measurements of $w$ in principle can be used to place limits on modified
gravity theories which attempt to explain cosmic acceleration without
dark energy \citep[e.g.,][]{2007PhRvD..76f3503W}. When self-consistent
models of non-linear collapse in such theories become available, it
sould be straightforward to use our cluster data in such tests also.

Comparison of the power spectrum normalization at $z=0$ obtained from
clusters with the amplitude of the CMB fluctuations is a sensitive
measure of the mass of light neutrinos. We constrain $\sum m_\nu <
0.33$~eV at 95\% CL, at the expense of slightly weakening the
measurement of $\wo$ obtained assuming that the neutrino masses are
negligibly small.

To facilitate the use of our cluster results in our cosmological studies, we
provide at the project WWW site\footnote{\texttt{http://hea-www.harvard.edu/400d/CCCP}} machine readable tables of the likelihood function computed on
several cosmological grids.

\acknowledgements

We thank Jeremy~Tinker for providing his mass function model results
prior to publication. We also thank W.~Hu, O.~Gnedin, M.~Markevitch, and
H.~Tananbaum for many useful discussions, and D.~Spergel and A.~Loeb for
comments on the manuscript. Financial support was provided by NASA
grants and contracts NAG5-9217, GO5-6120A, GO6-7119X, NAS8-39073 (AV,
WRF, CJ, SSM), GO5-6120C (HE); NSF grants AST-0239759 and AST-0507666,
NASA grant NAG5-13274 and the Kavli Institute for Cosmological Physics
at the University of Chicago (AK); Sherman Fairchild Foundation (DN);
FONDAP Centro de Astrofisica (HQ); Russian Foundation for Basic Research
grants RFFI 05-02-16540 and RFFI 08-02-00974 and the RAS program OFN-17
(RB and AV).

\bibliography{400d-cosm}

\end{document}